\documentclass[12pt,preprint]{aastex}

\begin{document}

\title{s-Process Nucleosynthesis in Advanced Burning Phases of Massive Stars}

\author{Lih-Sin The\altaffilmark{1},
Mounib F. El Eid\altaffilmark{2,1}, 
and 
Bradley S.  Meyer\altaffilmark{1}}
\altaffiltext{1}{Department of Physics and Astronomy,
Clemson University, Clemson, SC 29634-0978,\\ 
tlihsin@clemson.edu; mbradle@clemson.edu}
\altaffiltext{2}{Department of Physics, American University of
Beirut, Beirut, Lebanon; meid@aub.edu.lb}


\begin{abstract}
We present a detailed study of s-process nucleosynthesis in
massive stars of solar-like initial composition and masses 15, 20,
25, and 30 $M_{\sun}$. We update our previous results of
s-process nucleosynthesis during the core He-burning of these stars
and then focus on an analysis of the s-process under the physical conditions
encountered during the shell-carbon burning. We show that the
recent compilation of the $^{22}{\rm Ne}(\alpha,\rm n)^{25}Mg$ rate
leads to a remarkable reduction of the efficiency of the s-process during
core He-burning. In particular, this rate leads to the lowest overproduction
factor of $^{80}$Kr found to date
during core He-burning in massive stars.
The s-process yields
resulting from shell carbon burning turn out to be very sensitive
to the structural evolution of the carbon shell. This structure is influenced
by the mass fraction of $^{12}$C attained at the end of core helium burning,
which in turn
is mainly determined by the $^{12}{\rm C}(\alpha,\gamma)^{16}{\rm O}$ reaction.
The still present uncertainty in the rate for this
reaction implies that the s-process in
massive stars is also subject to this uncertainty. We identify some isotopes
like $^{70}$Zn and $^{87}$Rb as the signatures of the s-process during
shell carbon burning in massive stars.
In determining the relative contribution of our s-only stellar yields 
to the solar abundances, we find it is important to take into account the 
neutron exposure of shell carbon burning.
When we analyze our yields with a Salpeter Initial Mass Function,
we find that massive stars contribute at least 40\% to s-only nuclei
with mass A $\leq$ 87.
For s-only nuclei with mass A $>$ 90, massive stars contribute on average
$\sim$7\%, except for 
$^{152}$Gd, $^{187}$Os, and $^{198}$Hg
which are $\sim$14\%, $\sim$13\%, and $\sim$11\%, respectively. 

\end{abstract}




\section{Introduction}    \label{sec:intro}

The s-process nucleosynthesis is the slow neutron-capture process of
heavy nuclei in which the neutron-capture rate is slow relative to the
beta-decay rate of the heavy nuclei near the line of beta stability
\citep{1957RvMP...29..547B,1957PASP...69..201C}.
In this scenario,
the seeds to synthesize the heavy nuclei are the iron-group nuclei.
It is well known that massive stars of masses above M $\simeq$ 12$M_{\sun}$
are the sites of the so-called ``weak component'' of
s-process nucleosynthesis. The nuclei produced in this site are
restricted to the atomic mass range A $\simeq$ 65-90
\citep{1989RPPh...52..945K,1999ARA&A..37..239B}.
Many papers
have explored this weak component of the s-process, but
mainly during the core He-burning phase
\citep{1987ApJ...315..209P, 1989A&A...210..187L,
1990A&A...234..211P, 1991ApJ...371..665R, 1992A&A...258..357B,
2000A&A...354..740R, 2000ApJ...533..998T, 2004ApJ...611..396E}. 
Few papers
have investigated the s-process during the late evolution
phases of massive stars, especially during the phase of
shell carbon-burning (for short: shell C-burning) in these stars
\citep{1989A&A...210..187L, 1991ApJ...367..228R,
1993ApJ...419..207R}.
The study of the s-process during shell C-burning requires special
effort, because this burning episode lasts until the end of a
massive star's evolution, as indicated by many works dealing
with the advanced burning phases of massive stars
\citep{1988PhR...163...13N, 1998ApJ...502..737C,
2000ApJS..129..625L, 2002RvMP...74.1015W, 2004ApJ...611..396E}.

The contribution of the carbon shell to the s-process is
important because the nuclei synthesized in this site will be
eventually ejected largely unmodified during a supernova explosion
of the star due to the lack of a neutron source during the 
explosive burning. \citet{1993ApJ...419..207R}
argued on the basis of the calculations by \citet{1988PhR...163...13N}
that only stars of mass about 25 $M_{\sun}$
are able to eject the s-nuclei synthesized
in the C-shell. We further quantify this issue in \S \ref{sec:expdist}.
However, our main goal is to present a detailed study of the
s-process during core He-burning and shell C-burning, thereby
emphasizing the physical uncertainties influencing the results. We
will benefit from our detailed calculations
(\citealp[hereafter EMT04]{2004ApJ...611..396E}), where we have
investigated the influence of several important physical
quantities on the characteristics of the stellar models during the
advanced burning phases of massive stars.

In \S \ref{sec:smodels}, we summarize the main features of our
previous stellar evolution calculations, which we have carried
through the end of central oxygen burning. In \S \ref{sec:she}, we
present our updated results of the s-process during core
He-burning. In \S \ref{sec:spcshell}, we present the
characteristics of shell C-burning and discuss the
results obtained for the s-process during shell C-burning. 
In \S \ref{sec:expdist} we show the location 
(as function of the interior radius)
where nuclei are exposed to neutrons.
In this section, we also show the effectiveness of neutron captures
in producing heavy elements by calculating the total neutron capture
in the stellar models.
In \S \ref{sec:SolarComparison} we compare the production factor
distribution of s-only nuclei from our stellar models with the
distribution from the solar abundance.
In \S \ref{sec:summary} and
in \ref{sec:conclusion} we discuss and 
summarize the main conclusions of this
paper.

\section{Stellar Models}
\label{sec:smodels}

\begin{figure}
\epsscale{0.7}
 \rotatebox{90}{\includegraphics[height=6.5in]{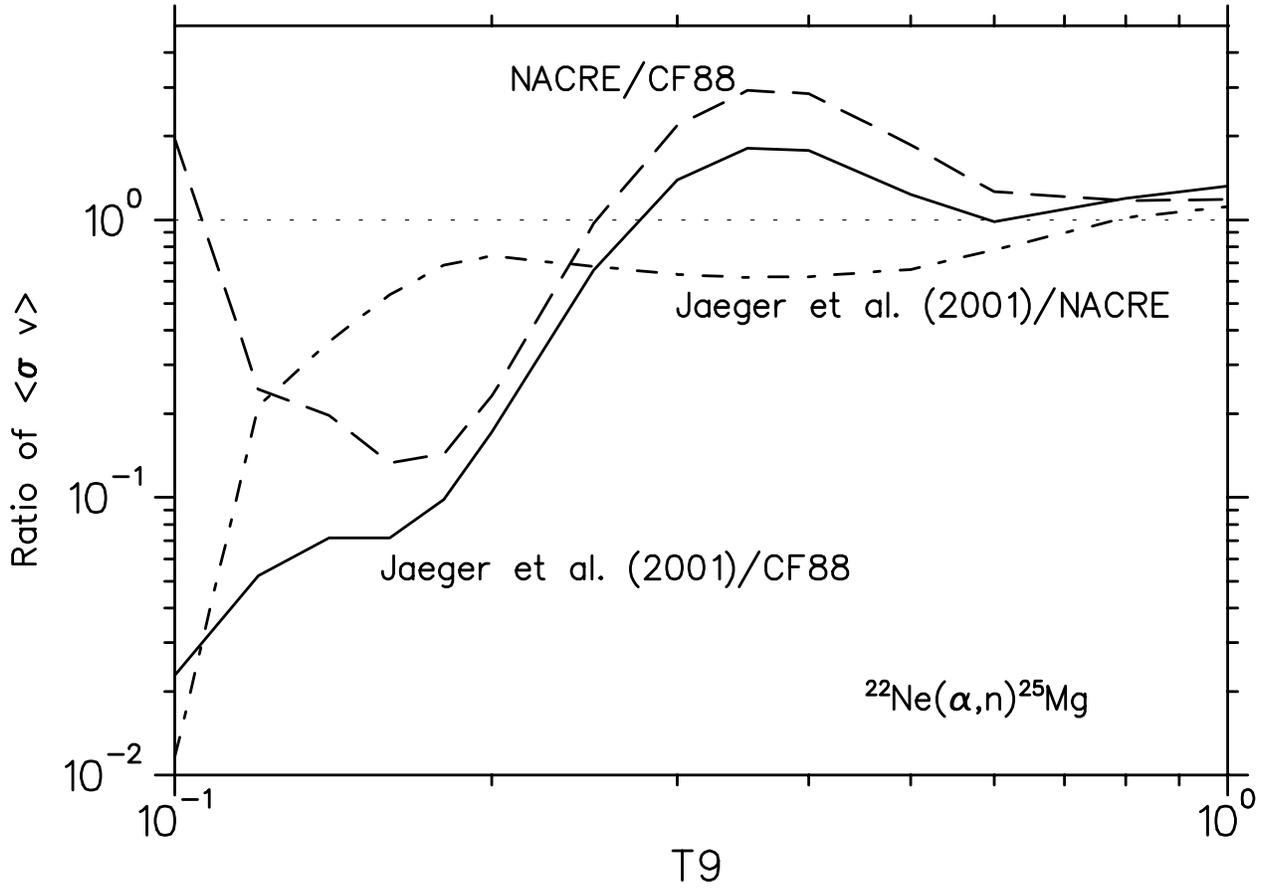}} 
 \caption{The
 \protect{$^{22}$Ne($\alpha$,{\rm n})$^{25}$Mg} rate 
 among different evaluations relative to CF88 or NACRE.
 See text for references.
 \label{fig:Ne22rate}
}
\end{figure}

The results of the s-process presented in this paper have been
obtained by using stellar models described in our previous
paper (EMT04) for stars of masses 15, 20, 25, and 30 $M_{\sun}$
with initial solar-like composition. We evolved our models
beyond core oxygen burning, and this allows us to investigate
the s-process nucleosynthesis not only during core He-burning, but
also during the important phase of shell C-burning. The network
we have used for the s-process 
is listed in Table 1 of                          
\citet[ hereafter TEM00]{2000ApJ...533..998T}.   
It includes 632 nuclei up to $^{210}$Bi and is sufficiently inclusive
that the s-process nucleosynthesis can be studied in detail.
The sources of the important nuclear reaction rates for each studied model
are summarized in Table \ref{tab:modelrate}.
The nuclear data have been updated as follows:
\begin{enumerate}
\item The nuclear masses were taken from the compilation by
\citet{1995NuPhA.595..409A}

\item The thermonuclear reaction rates were taken from the compilation
of the ``NACRE" collaboration \citep{1999NuPhA.656....3A}, and the
``Non-Smoker" rates according to \citet{2000ADNDT..75....1R}.
Most of the electron capture and $\beta$-decay rates (referred to as
weak interaction rates) are taken from
\citet{Nuclear..Wallet..Cards}. Certain weak interaction rates are
temperature and density dependent. These rates were taken from
\citet{1987ADNDT..36..375T}. However, we had to extrapolate some of
the weak interaction rates (e.g., the $\beta$-decay rate of
$^{79}$Se) to higher temperatures based on experimental
results by \citet{1988PhRvC..38..295K}. 
We have also used some of the extrapolated data
given by \citet{1993ApJ...419..207R} in their Table 2.

\item In our previous evolutionary calculations (EMT04), 
we investigated the effect of 
two different
$^{12}{\rm C}(\alpha,\gamma)^{16}{\rm O}$ 
reaction rates: 
the NACRE rate and that due to \citet{2002ApJ...567..643K}. 
The first rate is
larger than the second
in the temperature range T=(1-4)$\times$10$^8$ K
(Fig. 1 of EMT04),
which is relevant to
core He-burning in the massive stars under consideration. 
However, \citet{1996ApJ...468L.127B} recommends a rate that is
close to the rate given by \citet{2002ApJ...567..643K} at 
temperature T$_9$ $\leq$0.4 but is significantly larger at temperature 
range of 0.4 $\leq$ T$_9$ $\leq$ 3.0.
These
different rates lead to several consequences during the late stage
of core He-burning and also beyond this phase as
described by EMT04. Here, we summarize some of the consequences
that are relevant to the present study of the s-process.

In particular, using two different rates of the 
$^{12}{\rm C}(\alpha,\gamma)^{16}{\rm O}$
reaction in a 25 $M_{\sun}$ star of initial solar-like composition, we
found in the case labeled 25K in Table 1 of EMT04, 
where the rate is that according
to \citet{2002ApJ...567..643K}, that the mass fraction of carbon at the center
is X($^{12}$C)=0.280 (Table 4 of EMT04) at the end of core He-burning. 
In the case 25N, where the
NACRE rate was adopted, X($^{12}$C)=0.236. 
In case 25C, where we have used the rate 
by \citet[ hereafter CF88]{1988ADNDT..40..283C}, 
X($^{12}$C)=0.257. 
Finally, the case labeled 25NM, where mass loss has been neglected
during the evolution, has the lowest value X($^{12}$C)=0.193. This relatively
reduced value is due to the higher central temperature achieved during core
helium burning (see EMT04 for more details).

The lifetime of the
core carbon burning phase in Table 3 of EMT04 
is larger for a larger value of X($^{12}$C) as is the mass of the
convective core (indicated by $M_{\rm cc}$ in 
Table 3 of EMT04).
In addition, \S \ref{sec:spcshell}
shows that shell carbon-burning is sensitive to
the physical conditions achieved during core carbon-burning.  In
particular, it occurs in different regions of the star as indicated
by Figures 4 to 10 of EMT04.
A more detailed discussion of the effect of $X(^{12}{\rm C})$ 
is available in EMT04. In the present work, we
investigate its effect on the s-process itself. Our results are
presented in \S \ref{sec:she} and \S \ref{sec:spcshell}.

\item In our previous calculation of the s-process 
(TEM00), we adopted
a rate for the
neutron-capture reaction $^{16}{\rm O}({\rm n},\gamma)^{17}{\rm O}$
based on a cross section $\sigma_{16}$ =0.20 $\mu {\rm b}$ according to
\citet{1992ApJS...80..403B}. However, in our present calculations,
we have used the new rate for this reaction as obtained by
\citet{1995ApJ...441L..89I}. These authors have included the 434 keV
resonance and obtained a cross section $\sigma_{16}=34 ~\mu {\rm b}$
at T=30 keV or 170 times larger than the cross section obtained by
\citet{1992ApJS...80..403B}. The new rate is given by:

\begin{equation}
   <\sigma>_{16} = ({\rm kT})^{-1/2} + 5.88 ({\rm kT})^{1/2}
   \label{eq:sigma_16}
\end{equation}

Since $^{16}$O is a neutron sink, the new rate is expected
to reduce the efficiency of the s-process during core He-burning, as
has been emphasized by \citet{2000A&A...354..740R}. Our
results agree with this conclusion and are described in \S
\ref{sec:she}.

\item Another difference with our previous work (TEM00)
on the s-process concerns the reactions
$^{22}{\rm Ne}(\alpha,{\rm n})^{25}{\rm Mg}$ and
$^{22}{\rm Ne}(\alpha,\gamma)^{26}{\rm Mg}$.
The first reaction is known to be the main neutron source of the s-process in
massive stars, while the second is a competing reaction since it
captures part of the alpha particles.

Figure \ref{fig:Ne22rate} illustrates the rate of the
$^{22}{\rm Ne}(\alpha,{\rm n})^{25}{\rm Mg}$ reaction, 
where the NACRE rate is lower than that
obtained by CF88 in the
temperatures range below T$_8 \simeq\ 2.4$, a regime that comprises most
of the core He-burning phase in the stars under consideration. However,
the NACRE rate is larger by a factor of up to three in the temperature
range T$_8 \simeq\ (3-6)$,
which is relevant to the more advanced burning phases,
in particular the core and the shell carbon-burning.
Fig. \ref{fig:Ne22rate} displays the rate according to
\citet{2001PhRvL..87t2501J}, which shows some characteristics
similar to the NACRE rate,
although systematically slightly lower. 
Note that the situation at temperatures below
T$_8 \sim\ 2.0$ is controversial. 
Fortunately this temperature range does not affect
our results of the s-process.
In \S \ref{sec:she}, we show that the s-process yields do strongly
depend on the value
of the  $^{22}{\rm Ne}(\alpha,{\rm n})^{25}{\rm Mg}$ reaction rate.
\end{enumerate}

In our evolutionary calculations (EMT04), we have made a special
effort to analyze the effects of several important physical
quantities on the internal structure of the stellar models. Besides
considering the variation in $^{12}{\rm C}(\alpha,\gamma)^{16}{\rm O}$
as described above, we have investigated the effects of mass
loss on the structure of the stellar models.

Mass loss is certainly important for stars more massive than 15 $M_{\sun}$,
especially if the star evolves on the Kelvin-Helmholtz time scale to the
red giant stage, where mass loss is known to increase significantly
\citep{1988A&AS...72..259D}. As shown in our previous paper (EMT04), a rapid
evolution to the red giant branch occurs when the effect of the gradient of
molecular weight is taken into account. In this case, the Ledoux criterion for
convection inhibits the convective instability in the region of the
hydrogen-burning shell. Such evolution is found in the calculation
by \citet{2002RvMP...74.1015W} for the 15 and 25 $M_{\sun}$ stars.
We find that mass loss has an insignificant effect on the s-process during
core He-burning. However, as we shall see in \S \ref{sec:spcshell}, the
characteristics of shell-carbon burning are influenced, which may affect
the s-process in turn.

\section{s-Process in Core He-Burning: updated results}
\label{sec:she}

It is worth updating our previous results (TEM00) of the s-process during
core helium burning in massive stars, mainly because many reaction
rates determining the s-process efficiency have been revised as described in
\S \ref{sec:smodels}.

Table \ref{tab:hesprocess} presents a comparison of 
some key physical quantities characterizing
the efficiency of the s-process during core He-burning for the stars indicated.
Comparing our new results with the former ones, 
labeled TEM00 in Table \ref{tab:hesprocess},
we see that the new results show a significantly reduced 
s-process efficiency.
In particular, we obtain in the case of the 25 $M_{\sun}$ star (case 25C) 
an overproduction factor
of 618 for $^{80}$Kr compared to 1100 in our former calculations (TEM00). 
This remarkable decrease by a factor of 1.8 is mainly due to
the larger rate of
$^{16}{\rm O}({\rm n},\gamma)^{17}{\rm O}$ 
used in the present study.  Both the present calculation (25C) and that
in TEM00 used the same rate for the 
$^{22}{\rm Ne}(\alpha,{\rm n})^{25}{\rm Mg}$ reaction.
Therefore neutron capture on $^{16}$O is 
an efficient sink of neutrons, especially during
the advanced stage of core He-burning. It is interesting to
note that \citet{2000A&A...354..740R} 
found a reduction of 1.2 to 1.6 due to this
neutron sink. Hence, the statement in our previous work (TEM00) 
that most of the neutrons
captured by $^{16}$O will be returned by the reaction
$^{17}{\rm O}({\alpha},\rm n)^{20}{\rm Ne}$ is not fully justified
because it was
based on one-zone calculations in which the effect of convection was neglected.

In Tables \ref{tab:hesprocess} 
and \ref{tab:spresults} we show
the effect of the lower
NACRE rate for the $^{22}{\rm Ne}(\alpha,{\rm n})^{25}{\rm Mg}$ reaction
(see Fig. \ref{fig:Ne22rate}) as compared with 
that due to CF88 in the temperature
range up to T$_8 \simeq\ 2.4$.
It is clear that the NACRE rate leads to a significantly reduced 
s-process efficiency:
for example, Table \ref{tab:hesprocess}
shows that the overproduction of $^{80}$Kr in the case of the 25 $M_{\sun}$ 
star is 174 with the NACRE rate while it is 618 with the rate of CF88, 
a reduction by a
factor of 3.6. 
The low overproduction value is rather close to that obtained by
\citet{2000A&A...354..740R}. 
In their calculations for an 8 $M_{\sun}$ helium star
(corresponding roughly to an initial mass of 25 $M_{\sun}$),
they used the same rate as we do
for $^{16}{\rm O}({\rm n},\gamma)^{17}{\rm O}$ 
and considered different rates for
$^{12}{\rm C}(\alpha,\gamma)^{16}{\rm O}$.
They used the rate given by \citet{1993ApJ...414..735D}
for the
$^{22}$Ne($\alpha$,n)$^{25}$Mg, which is quite similar to the NACRE
rate shown in Fig. \ref{fig:Ne22rate}.
Taking the ``adopted'' NACRE rate for
$^{12}{\rm C}(\alpha,\gamma)^{16}{\rm O}$
(1.92 times larger than that of CF88 at T$_9$=0.3), 
they found an overproduction factor of 92 for $^{80}$Kr. 
However, when they used
the lower limit on the NACRE rate for
$^{12}{\rm C}(\alpha,\gamma)^{16}{\rm O}$
(1.16$\times$ of CF88)
the overproduction factor increased to 180.
The difference between their results and our 25N model
demonstrates the importance of 
the $^{22}{\rm Ne}(\alpha,{\rm n})^{25}{\rm Mg}$ rate.
Our results in Table \ref{tab:hesprocess} 
may be considered an update of the results of
\citet{2000A&A...354..740R}, 
since our rate for the 
$^{12}{\rm C}(\alpha,\gamma)^{16}{\rm O}$
is based on a new compilation as described in \S \ref{sec:smodels}.

One conclusion we may draw from the discussion above is 
that the efficiency of the
s-process during core He-burning in massive stars 
depends crucially on the neutron
production reaction 
$^{22}{\rm Ne}(\alpha,{\rm n})^{25}{\rm Mg}$ 
and on the neutron-sink reaction 
$^{16}{\rm O}({\rm n},\gamma)^{17}{\rm O}$.
Both reactions become effective during the late stage of core He-burning where
the $^{16}$O becomes abundant. Fortunately, the present
uncertainty in the 
$^{12}{\rm C}(\alpha,\gamma)^{16}{\rm O}$ 
does not have a significant
effect on the s-process efficiency,
as indicated by the comparison of the results
of the two cases ``present work, NACRE" and ``present work, 
(25K)" in which the rate for this
reaction is different as described in \S \ref{sec:smodels}.
Table \ref{tab:hesprocess} 
show that 
the efficiency of the s-process is remarkably
reduced for all the masses considered.

\begin{figure}
 \epsscale{0.7}
 \rotatebox{90}{\includegraphics[height=6.5in]{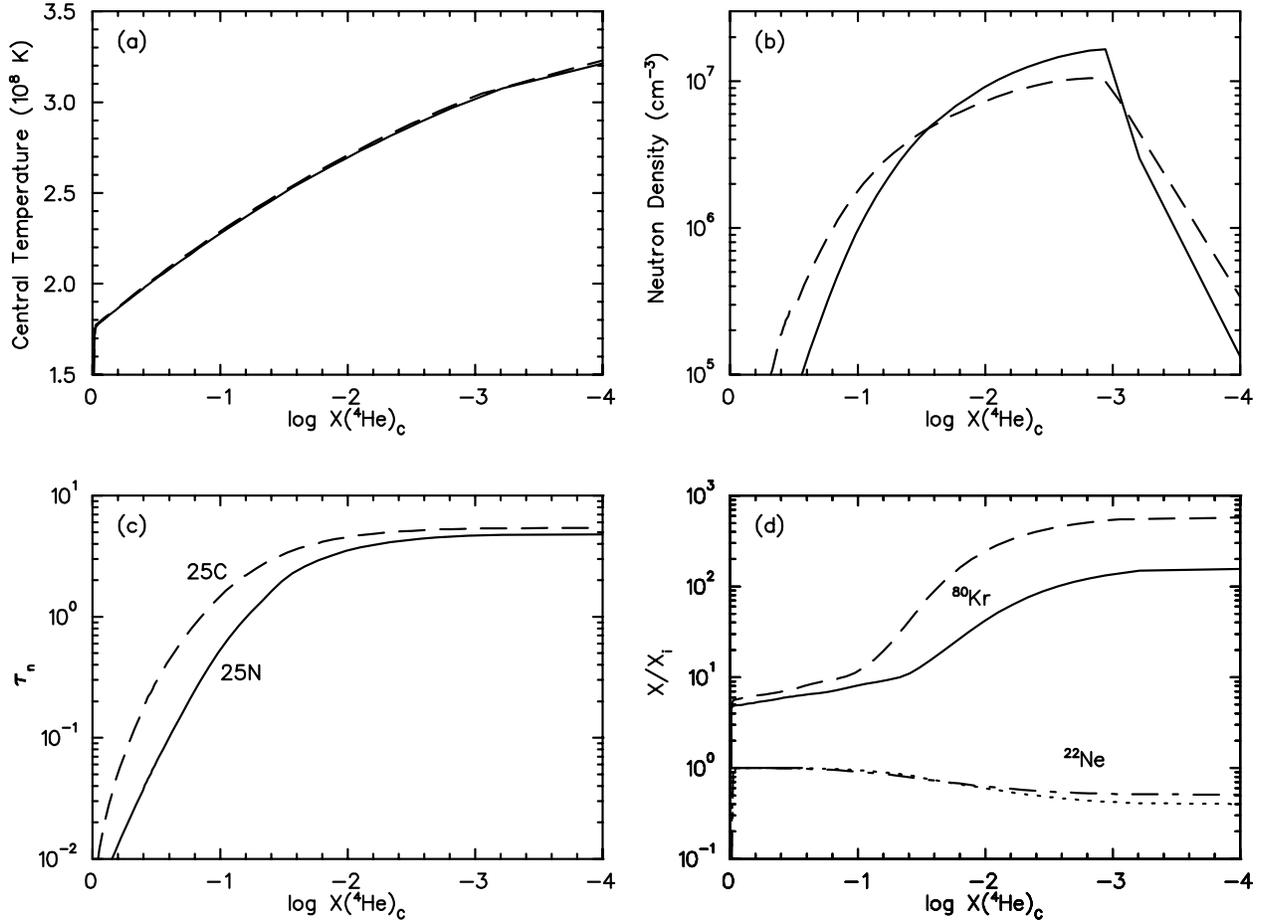}} 
 \caption{Some characteristics of the s-process during core
  He-burning according to the present work for 
  a 25 $M_{\sun}$ star in the cases 25N and 25C.
  $\tau_n$ is the neutron exposure experienced by a nucleus
  that remained at the center of the star at all times:
  $\tau_n = \int n_n \; v_{th} \; dt$, where
  $n_n$ and $v_{th}$ are the neutron density and thermal velocity
  of the neutrons at the center of the star.
  X/X$_i$ is the ratio of the mass fraction to the mass fraction at the
  beginning of core helium burning.
 \label{fig:corehe}
 }
 \end{figure}

It is worth analyzing our results of 
the s-process when the NACRE rate is applied for
the
$^{22}{\rm Ne}(\alpha,{\rm n})^{25}{\rm Mg}$ 
reaction instead of using the CF88 rate.
Fig. \ref{fig:corehe} displays several
physical quantities as a function of 
the central helium mass fraction as obtained for
the 25 $M_{\sun}$ models 25N and 25C during core He-burning. 
The larger rate by CF88 leads to 
an increase in the neutron density (see Fig. \ref{fig:corehe}b)
at larger helium mass fraction (or at earlier time).
Consequently, a higher neutron exposure,
$\tau_n(m_r) \equiv \int_0^t n_n(m_r,t') \; v_{\rm th} \; dt'$
\citep[Fig. \ref{fig:corehe}c]{1961AnPhy..12..331C}
is achieved leading to an earlier
increase of the overproduction of $^{80}$Kr (Fig. \ref{fig:corehe}d). 
In other words, despite the higher peak
neutron density achieved in case of the NACRE rate, 
the s-process is less robust because
the neutron exposure is lower due to
the shorter time scale until the end core He-burning.
The conclusion of this discussion is 
that the s-process during He-burning in massive stars is
a race against time since the whole process occurs 
during the late stage of this phase.

\begin{figure}
 \epsscale{0.7}
 \rotatebox{90}{\includegraphics[height=6.5in]{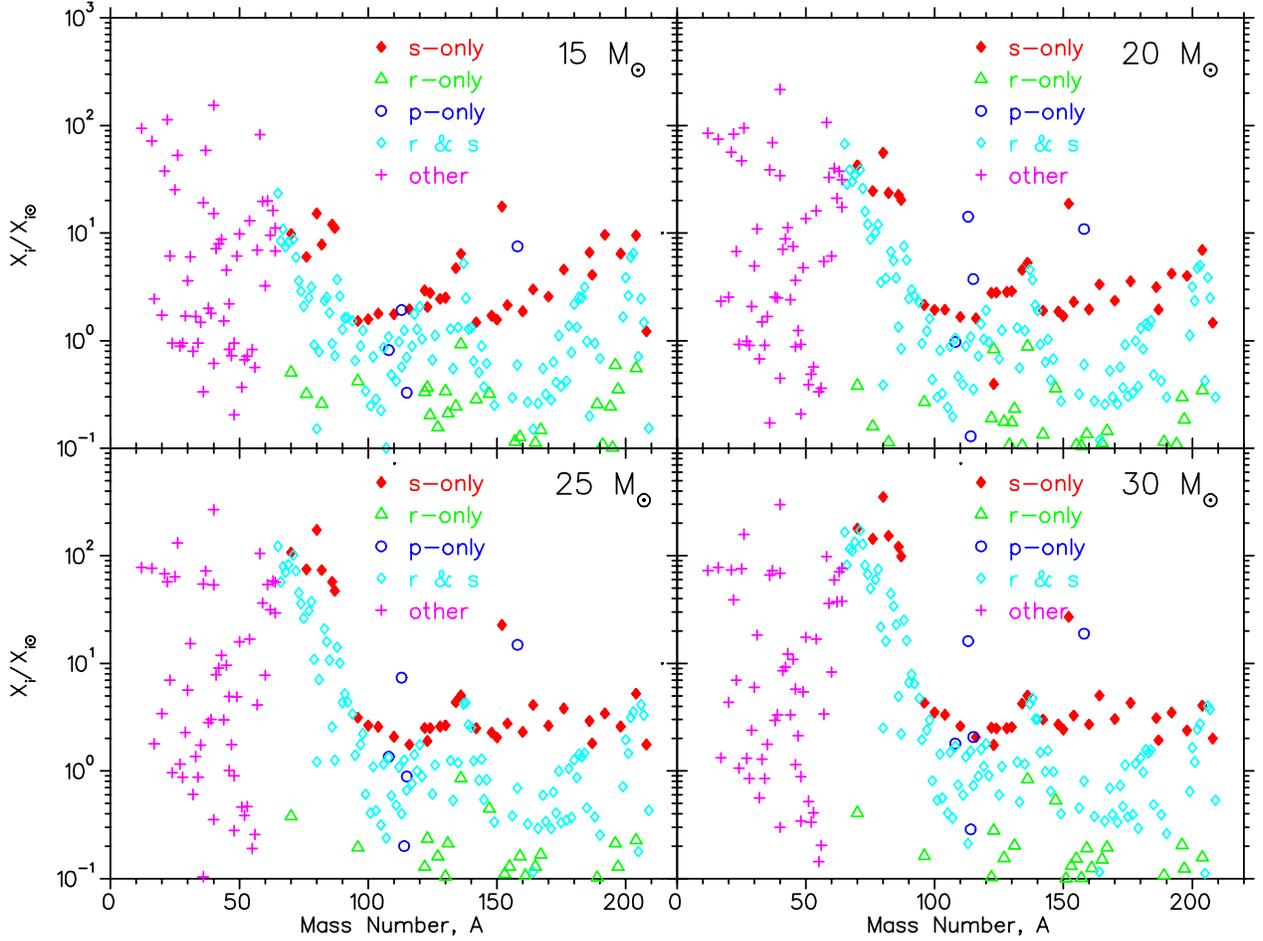}} 
 \caption{Overabundance factors of heavy nuclei averaged over
the convective helium burning core for model 15N, 20N, 25N, and 30N.
The primary nucleosynthesis production process for each nuclei 
is indicated by the symbol type.
 \label{fig:oCoreHe}
 }
\end{figure}

In Fig. \ref{fig:oCoreHe} we present the overabundances of heavy
nuclei averaged over the convective helium burning core for 
models 15N, 20N, 25N, and 30N, and 
in Tables \ref{tab:spresults} and \ref{tab:spsummary}
we list the overabundances of selected nuclei at the
end of the core helium burning.
The figures show that the larger the stellar mass, 
the more efficient is the neutron-capture of the s-process.
The figures also show the well-known feature of the weak s-process
that the overabundance distribution in the mass range A = 60 - 90
has a peak at $^{80}$Kr.

\section{s-process in shell carbon-burning}
\label{sec:spcshell}

At the end of core helium burning, the star begins to contract, and,
when its central temperature exceeds $\sim$5$\times$10$^8$ K, 
the neutrino energy loss dominates the energy balance 
\citep{2002RvMP...74.1015W}.
The carbon burning with its 
$^{12}$C + $^{12}$C reaction starts to be effective at temperature
$\sim$6$\times$10$^8$ K and density $\sim$3$\times$10$^4$ g cm$^{-3}$
(EMT04).
There are three effective reaction channels of $^{12}$C + $^{12}$C 
fusion and
overall carbon burning converts most of the initial carbon 
primarily into
$^{16}$O, $^{20}$Ne, $^{23}$Na, $^{24,25,26}$Mg, $^{28}$Si, and,
secondarily, into
$^{27}$Al and $^{29,30}$Si 
\citep{1983psen.book.....C,2002RvMP...74.1015W}.

In our previous evolution calculations (EMT04), 
we found that the carbon shell burning depends
sensitively on the profile of $^{12}$C resulting 
at the end of core carbon burning 
\citep{2002RvMP...74.1015W,2001ApJ...558..903I, 1996snai.book.....A}.
This profile
depends in turn on how core carbon burning proceeds: 
in a convective core or in a radiative region. 
The nature of that burning depends on the
central mass fraction of $^{12}$C attained at 
the end of core helium burning, which
is crucially influenced by the still uncertain 
$^{12}{\rm C}(\alpha,\gamma)^{16}{\rm O}$ reaction.
Thus, the rate for $^{12}{\rm C}(\alpha,\gamma)^{16}{\rm O}$
has a significant influence on the behavior of 
the shell C-burning, as we have previously shown (EMT04).
Our results of the s-process presented in the
following are expected to depend on this rate as well.

The neutron source of the s-process nucleosynthesis
during shell carbon burning
is the $^{22}$Ne($\alpha$,n)$^{25}$Mg reaction with 
an initial amount of X($^{22}$Ne)$\simeq$10$^{-2}$,
the amount left over from the end of core helium burning. 
The alpha particles for the $^{22}$Ne($\alpha$,n)$^{25}$Mg reaction 
are the product of the
$^{12}$C + $^{12}$C $\rightarrow$ $^{20}$Ne + $\alpha$
reaction channel.
The $^{22}$Ne($\alpha$,n)$^{25}$Mg cross section increases by a factor of
$\sim$10$^{10}$ from the phase of core helium burning (T$_9 \simeq$0.3)
to the phase of shell carbon burning (T$_9 \simeq$ 1.0-1.1).
The $^{22}$Ne ($^{4}$He) mass fraction during 
the s-process of core helium burning is 
$\sim$10$^{-1}$ ($\sim$10$^{-2}$),
whereas during the s-process of shell carbon burning,
X($^{22}$Ne)$\sim$10$^{-2}$ (X($^{4}$He)$\sim$10$^{-9}$).
These factors, together with the change in the density from
$\sim$10$^3$ g cm$^{-3}$ during core helium burning to
$\sim$10$^5$ g cm$^{-3}$ during shell carbon burning,
explain the difference in the neutron density during 
the s-process of core  helium burning (n$_n \sim$10$^7$ cm$^{-3}$) and
the s-process of shell carbon burning (n$_n \sim$10$^{10}$ cm$^{-3}$).

The consequence of the higher neutron density in the shell carbon burning
is the opening of the
(n,$\gamma$) path of some branchings in the s-process path.
Particularly important in the weak s-process are the branchings
at $^{79}$Se and $^{85}$Kr.
To illustrate the consequence, we take the data of $^{79}$Se at T=26 keV,
$< \sigma_{n,\gamma}>$= 225 mb and beta-decay t$_{1/2}$= 5.46 yr
and at T=91 keV, 
$< \sigma_{n,\gamma}>$= 97.3 mb and beta-decay t$_{1/2}$= 0.38 yr
\citep{1993ApJ...419..207R}.
The time scale of beta-decay is $\tau_{\beta} = t_{1/2}/ln(2)$ and
the time scale of neutron capture is $\tau_n = 1/n_n <\sigma v_{th}>$,
where the neutron thermal velocity 
$v_{th}$ = 2.4$\times$10$^8$ (T/30 keV)$^{1/2}$ cm s$^{-1}$.
Therefore when $\tau_{\beta} = \tau_n = \tau$, 
at T=26 keV, $n_n$ = 8.0$\times$10$^7$ cm$^{-3}$ and
at T=91 keV, $n_n$ = 1.4$\times$10$^9$ cm$^{-3}$.
This implies that during core helium (T$\simeq$26 keV) with
$n_n \simeq$ 10$^7$ cm$^{-3}$ ($<$ 8.0$\times$10$^7$ cm$^{-3}$)
most of $^{79}$Se beta-decays to 
$^{79}$Br(n,$\gamma$)$^{80}$Br($\beta^-$)$^{80}$Kr,
producing $^{80}$Kr.
However, during shell carbon burning (T$\simeq$91 keV), with
$n_n \simeq $10$^{10}$ cm$^{-3}$ ($>$ 1.4$\times$10$^9$ cm$^{-3}$),
the path of neutron capture $^{79}$Se(n,$\gamma$) is available and 
less $^{80}$Kr is produced.
The path differences followed by the s-process during these two burning
phases produce different ratios of $^{80}$Kr and $^{82}$Kr abundances.

\subsection{Shell Carbon Burning and Its s-process Characteristics}
\label{sec:cshcharact}

\begin{figure}
 \epsscale{0.7}
 \rotatebox{90}{\includegraphics[height=6.5in]{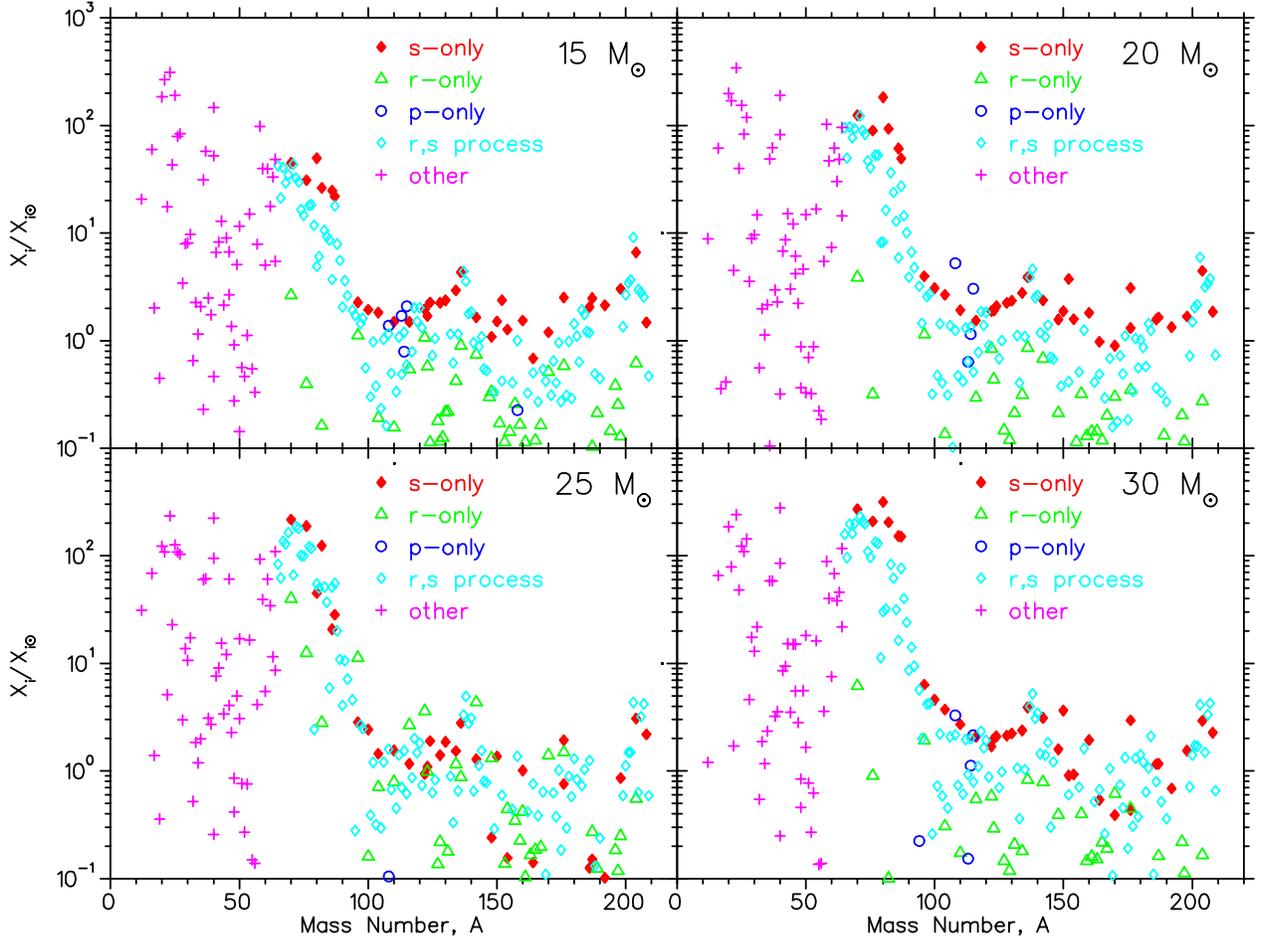}} 
 \caption{Similar to Fig. \protect\ref{fig:oCoreHe} but for 
  the convective carbon burning shell model 15N, 20N, 25N, and 30N.
 \label{fig:oCshell}
 }
 \end{figure}

Figures 4 to 10 of EMT04 show the convective structure of
our stellar models. Particularly relevant for this section are
the various locations of the carbon convective shells.
In Table \ref{tab:shellcarbon} we present some of the properties
of the last shell carbon burning phase in our stellar models.
In Figure \ref{fig:oCshell} we show the overabundance factors of
heavy nuclei averaged over
the convective carbon burning shell for model 15N, 20N, 25N, and 30N,
and in Tables \ref{tab:spresults} and \ref{tab:spsummary}
we tabulate some of their values.
The nucleosynthesis products from this carbon shell are the parts of
the carbon shells that would be ejected in the supernova event.
The locations, the mass ranges, and the durations of the C-shell
burning are not well correlated from one stellar model to the
others perhaps because of the variations due to the
mass fraction of $^{12}$C produced at the end of core helium burning,
to the temperature and density variations with the stellar mass, and
to the neutrino energy loss.
However, although the temperature and density of the innermost region
of the carbon burning shell vary quite significantly during the
convective carbon burning, their average values are quite similar
among the models. 
The average temperature at the bottom of convective shell is
$\sim$1.0-1.1$\times$10$^9$K and 
the average density is $\sim$1-2$\times$10$^5$ gm cm$^{-3}$.

The change of neutron exposures during shell C burning
$\Delta\tau_n$ at the bottom of the C-shell
convective zones (shown in Table \ref{tab:shellcarbon})
is significantly lower than the central $\tau_n$ produced
at the center during core helium burning 
(shown in Table \ref{tab:hesprocess})
by at least a factor of 4$\times$.
The average neutron exposures over convective zones $\langle \tau_n \rangle$ 
produced during C-shell burning
are also lower than during core helium burning by a factor of
2 in the 15 $M_{\sun}$ model and by a factor of 7 in 
the 30 $M_{\sun}$ model. 
The ratio of 
the average neutron capture per iron seed over the convective regions
$\langle \Delta n_c \rangle$
of core helium burning (Table \ref{tab:hesprocess})
relative to the value produced by carbon shell 
(Table \ref{tab:shellcarbon})
is $\sim$1  for the 15 $M_{\sun}$ model, 
increases to $\sim$2 for the 20 $M_{\sun}$ model,
and to $\sim$4-7 for the 25 $M_{\sun}$ models, and 
to $\sim$7 for the 30 $M_{\sun}$ model. This shows the trend of
increasing robustness of the s-process in the
carbon shell with decreasing stellar mass.

The maximum neutron density $n_n^{max}$ at the bottom of the
convective C-shell varies from 1.0$\times$10$^{10}$ cm$^{-3}$
to 70$\times$10$^{10}$ cm$^{-3}$ among our models, as shown
in Table \ref{tab:shellcarbon}.
The large variation is due to the fact that the
$^{22}$Ne($\alpha$,n)$^{25}$Mg rate varies by a factor of 17$\times$
for a 15\% change of temperature near T=1$\times$10$^9$K.
The difference in temperatures and densities at the bottom of the 
convective C-shell among models is about $\sim$20\%. 
These physical variables also vary during the C-shell evolution
as a result of the C-shell burning itself or
of the inner Ne and O shell burning.
Interestingly, the higher the temperature and density of the C-shell,
the shorter the burning duration.  This results in roughly the same
neutron exposure despite the higher neutron density.

The shell carbon burning decreases the mass fraction of $^{22}$Ne
by at least a factor of 5 from the value 
of X($^{4}$He)$\simeq$10$^{-2}$ 
at the end of core helium burning to 
X($^{4}$He)$\simeq$10$^{-3}$ at the end of
shell carbon burning.
This low value of X($^{22}$Ne) near the end of core oxygen burning
prevents significant change to
the heavy element abundances during the short time
left before the star explodes.
During the explosive phase, little subsequent alteration occurs
except for zones that achieve temperatures in
excess of T$_9 \approx 2.3$.
In such zones, the rate of $^{22}$Ne($\alpha$,n)$^{25}$Mg increases
by a factor of 3$\times$10$^4$ relative to the rate at T$_9$=1.0, but
the dynamic time scale decreases by a factor of 10$^7$
relative to the time scale during shell carbon burning.
Only disintegration reactions are likely to modify the abundances
in these zones significantly during the explosion
(e.g., \citealp{2003PhR...384....1A}).

The overabundance of $^{88}$Sr 
increases by a factor of $\sim$2
during shell carbon burning in each model studied,
and these overabundances increase
monotonically with increasing stellar mass.
In contrast to this, the overabundances of $^{80}$Kr decrease during
shell carbon burning for sequences 25N, 25NM, and 30N.
In these sequences, a significant fraction of the convective carbon
shell has a neutron density larger than $\sim$1.2$\times$10$^9$ n cm$^{-3}$.
At this neutron density (and at a temperature T$\approx 30$ keV),
the neutron-capture rate of $^{79}$Se dominates its beta-decay rate.
This allows the s-process flow to bypass $^{80}$Kr
thereby leading to its destruction (see also \citealp{1991ApJ...371..665R}).
We show this in more detail in the next section.

\subsection{25 $M_{\sun}$ s-process abundances}
\label{sec:25msunAbundance}

In this section,
we describe how the abundances of the s-process products change with time
as a massive star evolves through core He-burning and through shell C-burning.
We focus on the products of the s-process in the last shell C-burning
because these layers will be
ejected mostly without further nucleosynthesis processing
in a supernova explosion.

\begin{figure}
\epsscale{0.7}
 \rotatebox{90}{\includegraphics[height=6.5in]{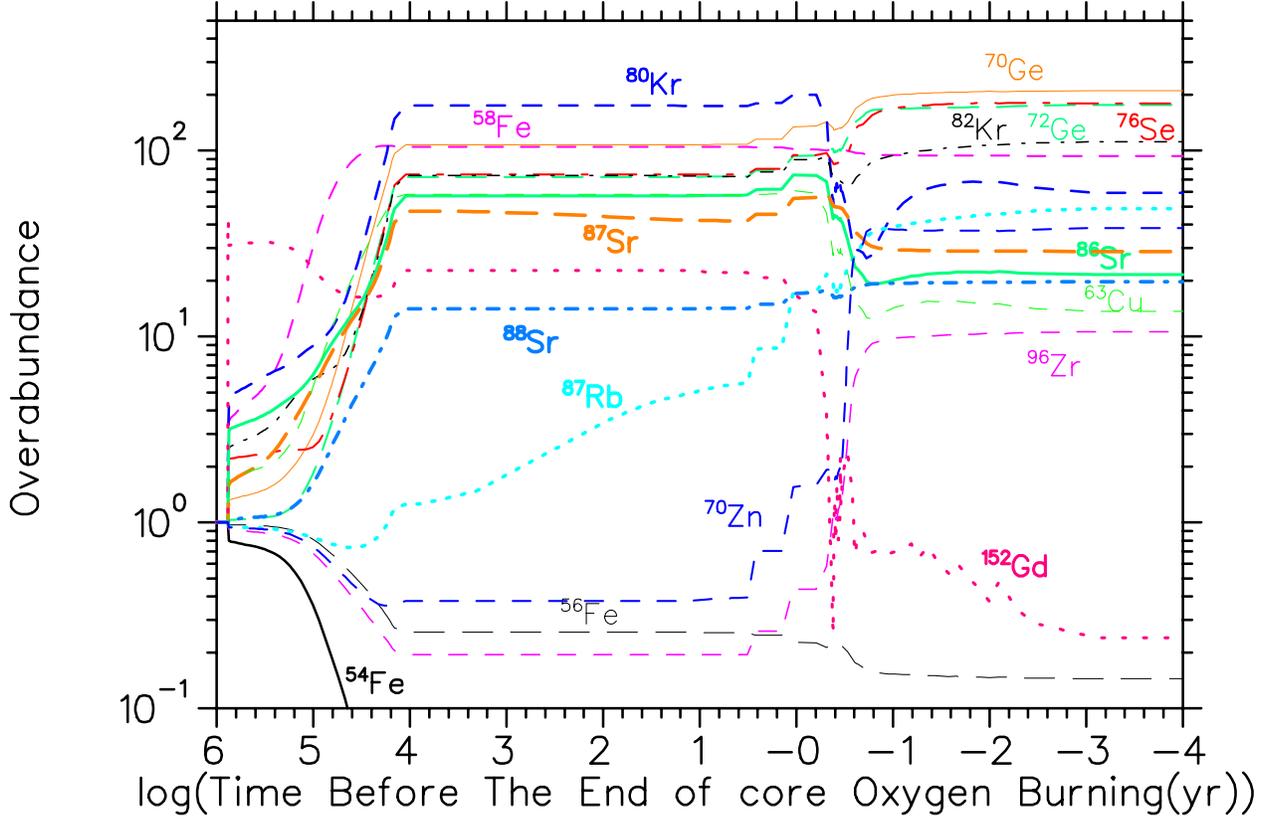}} 
 \caption{Overabundance factors of several nuclei 
  produced as a function of time by
 the s-process during core helium burning and shell carbon-burning
 in a 25 $M_{\sun}$ star, model 25N. 
  These factors are taken at mass coordinate of 2.26 $M_{\sun}$ 
  specifying the bottom of the convective carbon-burning
 shell in this case.  Core helium burning commences at abscissa value
 +6.0 and ends at +4.0.
 The first noticeable change
 occurs near the end of core helium burning, while the second change
 (abscissa value between 0.0 and -1.0) 
 is the result of shell carbon-burning. An exception is $^{87}$Rb
 which increases steadily between the two burning phases
 due to decay of $^{87}$Sr (see text for explanation).
 \label{fig:ab25n}
 }
 \end{figure}

In Fig. \ref{fig:ab25n} overabundance factors are shown as 
a function of time for the important
nuclear species produced by the s-process during core He-burning and 
shell C-burning in
a  25 $M_{\sun}$ star (evolutionary sequence labeled 25N in 
Table \ref{tab:modelrate}).
These curves represent the overabundance factors at 
a mass coordinate $M_{\rm r}$=2.26 $M_{\sun}$,
that is, at a mass shell
inside the convective core during core helium burning
but at the bottom of the convective
carbon-burning shell in this case (see Fig. 6 in EMT04).
The overabundances of all nuclear species shown in 
Fig. \ref{fig:ab25n} increase during
core He-burning except $^{54,56}$Fe, $^{70}$Zn, and 
$^{152}$Gd, which decrease because of
neutron capture. As expected, the pure s-nuclei
(${\rm ^{70}Ge, ^{76}Se, ^{80}Kr, ^{82}Kr, ^{86}Sr, ^{87}Sr}$) 
are produced in particular
during this phase, as summarized in Table \ref{tab:spresults}.

The modification of the s-process products by 
shell C-burning is most effective during core
neon burning and core oxygen burning in case 25N,
as shown in Fig. \ref{fig:ab25n} at the time
coordinate between 0.0 and -1.0. 
In this case, the convective carbon-burning shell is
most effective as it has settled in 
the mass range 2.26-4.94 $M_{\sun}$ (see Table \ref{tab:shellcarbon}
and Fig. \ref{fig:taun}). 
Note that the increase of the overabundance of $^{87}$Rb
before the onset of shell C-burning is a result of the 
$\beta^+$-decay of $^{87}$Sr, whose rate
is sensitive to temperature according to the work 
of \citet{1987ADNDT..36..375T}.

\begin{figure}
\epsscale{0.7}
   \rotatebox{90}{\includegraphics[height=6.5in]{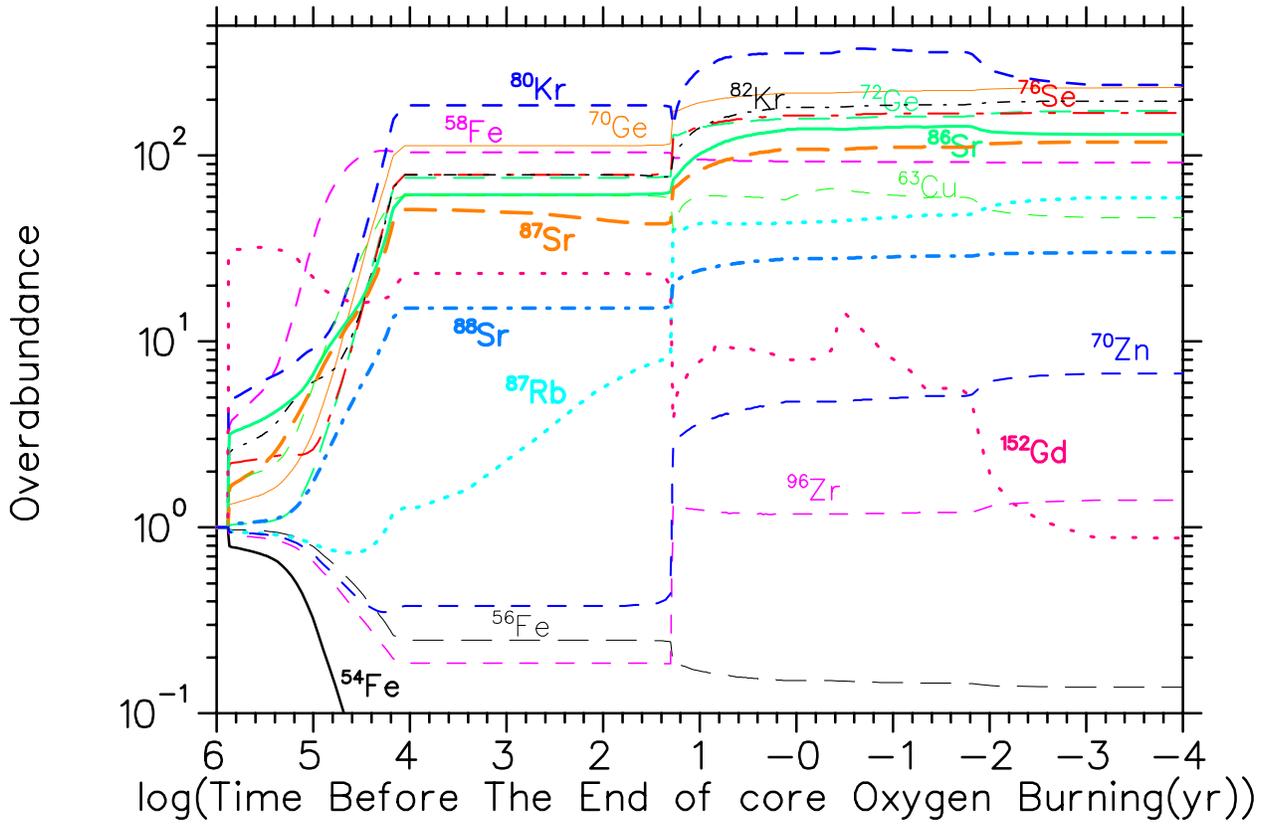}} 
   \caption{The same as Fig.4 but for the case 25K in our
   calculations. The overabundance values are determined at mass
   coordinate 2.0 $M_{\sun}$ in this case.
   \label{fig:ab25k}
   }
   \end{figure}

\begin{figure}
\epsscale{0.7}
  \rotatebox{90}{\includegraphics[height=6.5in]{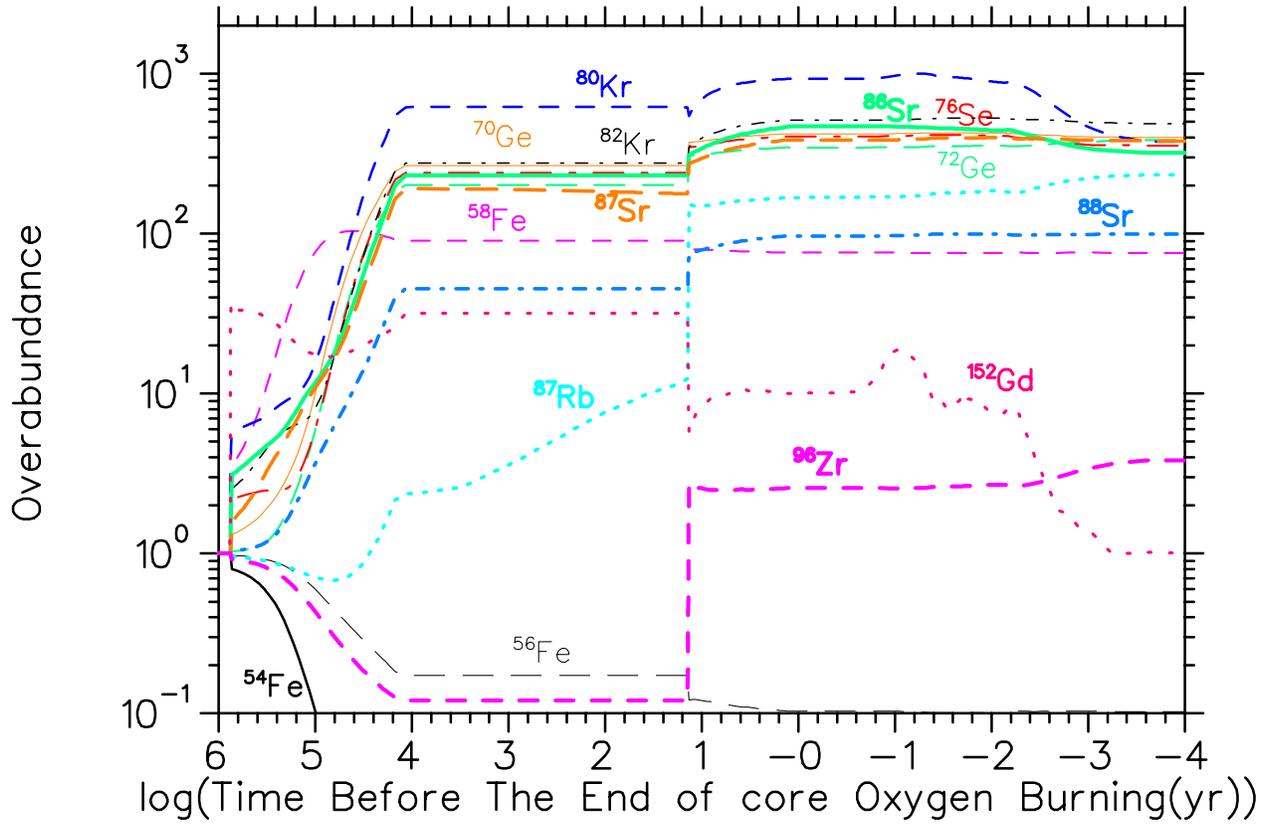}} 
  \caption{ The same as Fig.4 but for the case 25C in our
  calculations. The overabundance values are taken at mass
  coordinate 2.547 $M_{\sun}$.
  \label{fig:ab25c}
  }
\end{figure}

Figs. \ref{fig:ab25n}, \ref{fig:ab25k}, and \ref{fig:ab25c}
show that the effect of 
shell C-burning on the overabundance
factors is distinct from that due to 
core He-burning (time coordinate between 4.0 and 5.0)
for case 25N, 25K, and 25C, respectively.
The neutron density achieved in case 25 N during 
shell C burning achieves a peak value of
$\simeq 7\times10^{11} \rm {cm}^{-3}$ 
(see Fig. \ref{fig:ndin25n}). Therefore,
the branchings at the sites of the unstable nuclei 
$^{63}$Ni, $^{69}$Zn, $^{79}$Se, and $^{85}$Kr
become effective. This can be traced in 
Fig. \ref{fig:ab25n} and Table \ref{tab:spresults}
as follows:
\begin{itemize}
 \item  The decrease of the overabundance of
        $^{63}$Cu and $^{65}$Cu and the increase of $^{64}$Ni 
        indicate the branching at $^{63}$Ni.
 \item  The branching at $^{69}$Zn is indicated by 
        the increase of the overabundance of $^{70}$Zn
        and the decrease of the abundance ratio of the Germanium isotopes
        $^{70}$Ge/$^{72}$Ge.
        Notice that $^{70}$Zn was destroyed during core He-burning, 
        but produced by shell C-burning 
        (see \ref{sec:zn70} for more discussion on $^{70}$Zn).
 \item  The branching at $^{79}$Se leads to a modification 
        of the overabundance of the isotopes
        $^{80,82}$Kr. However, 
        the overabundance of $^{80}$Kr is diminished by a factor of
        about 4 compared to its value reached at 
        the end of core He-burning while
        the overabundance of $^{82}$Kr increases by a factor of about 1.7.
\item   The effect of the branching at $^{85}$Kr leads to the increase
        of the overabundance of the isotopes $^{87}$Rb 
        \citep{1993ApJ...419..207R}.
\item   Finally, there is a decrease in the abundance of the
        isotopes $^{86,87}$Sr (affected by the branchings at 
        $^{85}$Kr and $^{86}$Rb)
        and an increase 
        of the overabundance of $^{96}$Zr to a value larger than 1
        ($^{96}$Zr is destroyed during core helium burning).
\end{itemize}

In the cases 25K, the second stage of convective shell carbon-burning
comprises an extended mass range of 1.30-4.54 $M_{\sun}$ 
(see Table \ref{tab:shellcarbon}).
The overabundance factors are shown in Fig. \ref{fig:ab25k} 
and are taken at a mass
coordinate of 1.38 $M_{\sun}$ specifying the bottom 
of the convective carbon-burning
shell in this case 
with its physical condition as a function of time 
shown in Fig. \ref{fig:ndin25k}.
These factors are distinct from those in the case
25N in many respects:
\begin{itemize}
 \item The overabundance of $^{80}$Kr is increased 
       by shell C-burning in contrast to the case
       in 25N.
       The reason is the lower neutron density achieved during this phase in
       this sequence (see Fig. \ref{fig:ndin25k}).
 \item This lower neutron density explains the
       lower overabundance factor of $^{86}$Kr 
       (about a factor 5 lower than in case 25N)
       and also the relatively higher overabundance of $^{86,87}$Sr. 
       In other words, the
       nuclear-reaction flow in case 25K 
       does not proceed beyond the Sr isotopes
       to reach the Zirconium region, as in case of 25N.
\end{itemize}

In the case of 25C the overabundance factors are 
displayed in Fig. \ref{fig:ab25c}. Despite the
similarity with Fig. \ref{fig:ab25k}, 
the s-process is generally more efficient in this
sequence in both the core He-burning or the shell C-burning
due to the adoption of the CF88 rates.

It is worth relating these differences to 
the evolution of the stellar models.  To do so,
we compare Figures \ref{fig:ndin25n} and \ref{fig:ndin25k}, 
which show several physical variables
characterizing the properties of the carbon-burning shell 
in the cases 25N and 25K, respectively.
We recall that 25N 
at the end of core helium burning
has a central carbon mass fraction X($^{12}$C)=0.236 
while the 25K has
X($^{12}$C)=0.280. 
This slight difference has a significant effect on the ensuing evolution
beyond core He-burning, as described in detail by EMT04.

We emphasize some points that help to understand 
the s-process during shell C-burning. 
The shell burning proceeds differently in the sequence 25N and 25K.
The relatively lower value of X($^{12}$C) of the sequence 25N 
leads to a convective core of mass  
$M_{CC}$=0.36 $M_{\sun}$, 
while $M_{CC}$=0.47 $M_{\sun}$ in the case 25K.
Due to this difference, shell C-burning proceeds in 25N in three
convective episodes but in two convective episodes in 25K.
This can be seen in Figs. \ref{fig:ndin25n} and \ref{fig:ndin25k},
especially in the behaviour of the neutron number density as a function
of time.
A peak value of the neutron density of $\sim$7x10$^{11}$ cm$^{-3}$ 
is achieved in 25N compared to $\sim$1.1x10$^{10}$ cm$^{-3}$ in case 25K.
The higher neutron density in 25N creates a flow of neutron-capture 
reactions which reaches the region of Zirconium 
(see Fig. \ref{fig:cpath}.
On the other hand, the third convective C-shell phase in 25N lasts 
$\sim$0.5 yr,
while the second convective C-shell in 25K lasts for $\sim$24 yr, 
or $\sim$50$\times$ longer.
This longer burning time leads to more depletion of carbon in the 
C-shells where X$_f(^{12}$C)=7.2$\times$10$^{-3}$ in 25K
compared to X$_f(^{12}$C)=9.4$\times$10$^{-2}$  in 25N.

The neutron density and the duration of neutron exposure in the two models
produce the neutron exposure, 
$\tau_n \simeq$0.9 mbarn$^{-1}$
at the bottom of the respective convective carbon shells.
However, the longer duration and larger mass range of 
convective carbon shell burning in 25K ($M_r$ = 1.30-4.54 $M_{\bigodot}$) 
than in 25N produces
a higher value of neutron capture per iron seed nucleus $n_{cap}$
in 25K (see Figs. \ref{fig:ndin25n} and \ref{fig:ndin25k}).
In order to explain those values of $\tau_n$ and $n_{cap}$,
we performed several test calculations of nuclear burning and mixing
of the convective zones.  

As a reference calculation, 
we constructed spherical shells with
the temperatures, densities,
diffusion coefficients, and mass coordinates of model 25K when
the convective C-shell is at its maximum extent
($M_r$ = 1.30-4.54 $M_{\bigodot}$).
As the initial composition of all test calculations of C-shell burning, 
we took the composition of the stellar model at the end of
core helium burning, at which point
$n_{cap}$=3.70.
We ran a simultaneous nuclear burning and mixing code \citep{openIssues}
for about 20 yrs duration
so that X$_f(^{12}$C)$\leq$1$\times$10$^{-3}$.
In this simultaneous burning and mixing code, the temperature, density,
and diffusion coefficient of each zone remained fixed in time for simplicity.
The results of this reference calculation show
a neutron exposure at the bottom of the convective shell of
$\tau_n$=0.20 mbarn$^{-1}$ and a number of neutrons captured per Fe seed of
$n_{cap}$=4.74.

To understand the effect of the size of convective zones
(since model 25N has a thinner range of convective zones than in model 25K),
we ran a second calculation in which we
reduced the thickness of convective zones
to about 70\% in mass range.  We kept the other variables
the same as in the reference calculation. 
In this second calculation, we obtained
$\tau_n$=0.07 mbarn$^{-1}$ and  $n_{cap}$=4.74 
by the time X$_f(^{12}$C) had dropped to $1 \times 10^{-3}$ in about
20 years.  This shows that the less available neutron source due to
the reduced size of convective zones produces a smaller
value of neutron exposure. On the other hand, there were a correspondingly
smaller number of seed nuclei, so the number of neutron captures per
Fe seed nuclei was the same as in the reference calculation.
The relation between this test calculation to the reference one
clearly does not mimic the relation between the 25N and 25K results.

To test the effect of the higher temperature in the carbon shell in 25N,
we performed another test calculation in which
we kept all physical variables the same
as in the reference calculation but increased the temperature of each zone
by 15\%.
For this structure, we found
$\tau_n$=0.21 mbarn$^{-1}$ and  $n_{cap}$=3.85 at time t=0.04 yr
and
$\tau_n$=0.26 mbarn$^{-1}$ and  $n_{cap}$=4.86 at time t=20 yr.
This test shows that the structure with 
higher temperature produces a much higher
neutron density and hence the same value of $\tau_n$
($\approx 0.2$ mbarn$^{-1}$)
as in the reference calculation but in only 0.04 years.
Because of the shorter time to reach $\tau_n = 0.21$ mbarn$^{-1}$,
this test calculation
yields a smaller value of $n_{cap}$ at the same $\tau_n$.
This result indeed mimics the difference between the results
of 25N and 25K, and we
conclude that the similar values of $\tau_n$ in the
C-shell of 25N and 25K and the higher value of $n_{cap}$
in 25K at the end of their C-shell burning 
is mostly due to the higher temperature in 25N than in 25K.

In Figure \ref{fig:finalx} we show the overabundance 
distribution of 
$^{12}$C, $^{16}$O, $^{22}$Ne, $^{70}$Zn, $^{70}$Ge,
$^{80}$Kr, and $^{86}$Sr--nuclei that are important for 
the s-process nucleosynthesis during
shell carbon burning--at the end of core oxygen burning
for sequences 25N and 25K to illustrate
Table \ref{tab:spresults} and the discussion above.
The values of the overabundances of light nuclei 
$^{12}$C, $^{16}$O, and $^{22}$Ne
drop significantly from 92(78), 72(77), 60(59)
at the end of core helium burning to
2.4(31), 61(68), and 3(5) at the end
of core oxygen burning for the last carbon shell region
in sequence 25K(25N), respectively.
The overabundance of the s-only nuclide $^{70}$Ge increases
by a factor of $\sim$2$\times$ for both sequences during 
shell carbon burning. 
The overabundance of $^{70}$Zn at the end of core helium
burning is 0.4 for sequences 25N and 25K, but increases
significantly to a value of 5 for sequence 25K and
to a value of 40 for sequence 25N
during shell carbon burning, as shown in the figure.

Our general conclusion of this analysis is that s-process nucleosynthesis 
occurring in shell C-burning
is rather sensitive to the central mass fraction of $^{12}$C 
attained at the end of core He-burning,
and this in turn depends on the rate of 
$^{12}{\rm C}(\alpha,\gamma)^{16}{\rm O}$, a value still
under debate, as outlined in \S \ref{sec:smodels}.

\subsubsection{$^{70}$Zn}
\label{sec:zn70}

The production of $^{70}$Zn in explosive carbon and 
neon burning is discussed in \citet{1996snai.book.....A}.
The source of neutrons for the neutron-capture reactions 
in explosive burning is the $^{12}$C+$^{12}$C reaction. 
In C or Ne explosive burning, $^{70}$Zn
is produced in almost equal amounts as $^{68}$Zn
\citep{1996snai.book.....A, 1972ApJ...175..201H}
whereas the ratio of $^{70}$Zn to $^{68}$Zn in solar abundance is 0.033.
An analysis by \citet{1996snai.book.....A}
concludes that solar $^{70}$Zn is produced in a nuclear
burning process with time scale that is longer than a typical
explosive time scale, which suggests
the hydrostatic burning of carbon or neon as the site
production of $^{70}$Zn.  In such burning, the $^{70}$Zn
overproduction should be a fraction of $^{68}$Zn overproduction, as
we find in the present analysis.

The solar abundance of $^{70}$Zn isotope is 0.62\% of all Zinc isotopes
\citep{1989GeCoA..53..197A}.
The small fraction of $^{70}$Zn abundance  relative to other Zinc
isotopes makes it difficult to detect in the spectra
of stellar atmosphere or interstellar medium.   It is possible that
hints of $^{70}$Zn might be preserved in meteoritic samples.
For example, $^{70}$Zn isotopic anomalies have been measured in
Allende meteorite inclusions.
A clear excess of $^{66}$Zn 
and a deficit of $^{70}$Zn
in FUN inclusions
\citep{1990ApJ...358L..29V, 1990ApJ...360L..59L} 
is correlated
with excesses for the neutron-rich isotopes of 
$^{48}$Ca, $^{50}$Ti, $^{54}$Cr, and $^{58}$Fe.
The source of these anomalies was attributed to neutron-rich e-process
nucleosynthesis in massive stars \citep{1985ApJ...297..837H}, but the
current thinking is that these isotopes were produced in rare Type Ia
supernova (e.g., \citealp{1996ApJ...462..825M,1997ApJ...476..801W}).
Since such nucleosynthesis does not produce $^{70}$Zn, the correlated
deficit of this isotope with excesses in $^{48}$Ca, for example, is
expected.  A more promising cosmochemical sample that might provide
evidence of $^{70}$Zn production in C-shell s-processing is a presolar
grain.
It is quite reasonable to expect that some shell carbon burning
products might condense or be implanted into grains which,
if then preserved in meteorites, would show the excesses of
$^{70}$Zn and $^{87}$Rb isotopes, along with other s-process products listed in
Table \ref{tab:spresults} for correlation analysis.

\subsubsection{$^{87}$Rb}
\label{sec:rb87}
Rb solar abundance is comprised of $^{85}$Rb and $^{87}$Rb isotopes with
the $^{87}$Rb abundance 27.8\% of the total \citep{1989GeCoA..53..197A}.
The abundance of the long-lived radioactive $^{87}$Rb, which decays to
daughter nuclei $^{87}$Sr with a half-life of 4.9$\times$10$^9$ yr,
is often used in radioactive dating of rocks and meteorites
(e.g., \citealp{1995aitc.book.....C, 2006E&PSL.246...90M}).
It is probably not possible to observe Rb isotope ratio directly
from massive stars, however, the ejected abundances into interstellar medium
or molecular clouds could be measured.
Recently \citet{2004ApJ...603L.105F} reported
the first measurement of the interstellar $^{85}$Rb/$^{87}$Rb isotope ratio
from the diffuse gas toward $\rho$ Oph A.
They obtained a value of 1.21 which is
significantly lower than the solar abundance value of 2.59.
A proper understanding of the origin of $^{87}$Rb in the diffuse gas
will require
chemical evolution calculations with mixing of several generations of
stars. Correlating the $^{87}$Rb observed abundance with
other heavy isotopic abundances
could reveal interesting insights into carbon-shell nucleosynthesis.

\subsection{s-process path}
\label{sec:sprocpath}

In order to see the differences in the s-process paths of 
core helium burning and 
shell carbon burning, we performed two one-zone
s-process nucleosynthesis calculations using the
central temperature, central density, central mass fraction
X($^{4}$He) of the core helium burning in sequence 25N and
the temperature, density, and mass fraction X($^{12}$C) of
the innermost shell of the shell carbon burning in sequence 25N,
respectively. 
The initial composition of the calculation is taken from
the initial composition of the burning phase in sequence 25N.
For each time step, we compute the reaction flow.  For example,
for the reaction $i + j \to k + l$, the
integrated flow over time step $dt$ is
$f_{i+j,k+l}$ = N$_A<\sigma v>_{ij,kl} \rho$ Y$_i$ Y$_j$ dt,
where $Y_i$ and $Y_j$ are the abundances of species $i$ and
$j$, respectively.
The total of the integrated net flow for all timesteps,
$F_{i+j,k+l} = \sum_{n} [f_{i+j,k+l} - f_{k+l,i+j}] \; dt_n$
shows the
total flow of the reaction.
The net integrated nuclear reaction flows are shown for the case 25N in 
Fig. \ref{fig:hepath} and 
Fig. \ref{fig:cpath} for the s-process
during core He-burning and shell C-burning, respectively. 
Similar features shown in Fig. \ref{fig:hepath} and
Fig. \ref{fig:cpath} are also reproduced in sequence 25K.
The differences in temperature, density, and, therefore, the neutron density,
cause some differences in the branchings of the flows.
The (n,$\gamma$) branching paths that are 
opened or enhanced during the third convective
C-shell relative to s-process paths during the core helium burning in 25N 
are at 
$^{57,60}$Fe, $^{64}$Cu, $^{69}$Zn,
$^{70}$Ga, $^{75}$Ge, $^{76}$As, $^{81}$Se, 
$^{80,82}$Br, $^{85}$Kr, $^{86,87}$Rb, 
$^{87,89,90}$Sr, $^{90}$Y, and $^{95}$Zr.

It is interesting to note that Fig. \ref{fig:cpath} shows 
that proton-capture reactions occur on nuclei with Z=26-30 
during shell C-burning.
The largest (p,$\gamma$) flow for Z$\geq$26 is the
$^{58}$Fe(p,$\gamma$) flow. The ratio of
$^{58}$Fe(p,$\gamma$)/$^{58}$Fe(n,$\gamma$)
$\simeq$ 1.4$\times$10$^{-3}$, which clearly shows that proton capture
reactions do not affect the s-process flow.

In these one-zone nucleosynthesis calculations of shell
carbon burning, we find the five largest neutron source
reactions are the ($\alpha$,n) reactions on
$^{22}$Ne, $^{21}$Ne, $^{17}$O, $^{13}$C, and $^{26}$Mg,
with $^{21}$Ne produced through $^{20}$Ne(n,$\gamma$)$^{21}$Ne,
$^{17}$O produced through $^{16}$O(n,$\gamma$)$^{17}$O,
$^{13}$C produced through $^{12}$C(p,g)$^{13}$N($\beta^+$)$^{13}$C,
and $^{26}$Mg produced through 
$^{12}$C + $^{12}$C $\rightarrow$ $^{23}$Na + p followed by
$^{23}$Na($\alpha$,p)$^{26}$Mg. 
The alphas (58\%) and protons (42\%) 
are produced by the two channels of carbon burning.

Recently \citet{2004ApJ...601..864T} in their study of 
Sr, Y, and Zr Galactic evolution infer some hints of 
a primary s-process in low-metallicity massive stars.
These authors suggested that 
$^{12}$C + $^{12}$C $\rightarrow$ $^{23}$Mg + n 
and $^{26}$Mg($\alpha$,n)$^{29}$Si  
during carbon burning
could be the neutron source reactions in the extremely
metal poor (EMP) massive stars. 
To analyze this suggestion, we evolved a 25 $M_{\sun}$ star
with initial metallicity [O/H]=[Fe/H]= -4 up to the end of 
core helium burning and then took this composition 
as the initial composition of a one-zone calculation of
shell carbon burning using the physical conditions of
sequence 25N above.
We find the ratios of $\alpha$, p, and n channels of
$^{12}$C + $^{12}$C reaction flow are 1.0:0.71:0.0, respectively.
The largest neutron source reactions are the ($\alpha$,n)
reactions on $^{13}$C, $^{17}$O, $^{21}$Ne, and $^{22}$Ne
with their ratios of 1.0:0.17:0.06:0.02, respectively.
The $^{13}$C($\alpha$,n)$^{16}$O
reaction is the major neutron source during shell carbon
burning, which can also be shown for core helium burning 
\citep{1992A&A...258..357B, 2000A&A...354..740R}, instead of
the $^{22}$Ne($\alpha$,n)$^{25}$Mg reaction in the EMP
massive star.  We find that most of the Sr nuclei (77\%) are produced
in core helium burning rather than in the shell carbon
burning. Therefore, carbon burning could not provide 
enough neutrons to explain the enhancement of
the observed Sr abundances in the EMP stars.

\section{Neutron Exposure and Neutron Capture per Iron Seed Nuclei}
\label{sec:expdist}

Several burning sequences in massive stars produce neutrons 
through the $^{22}$Ne($\alpha$,n)$^{25}$Mg reaction.
A simple way to show where nuclei are exposed to these neutrons
during the stellar lifetime is to plot the
total neutron exposure versus the interior mass radius as shown
in Fig. \ref{fig:taun} 
at the end of core oxygen burning
(the last model calculated).
$\tau_n(M_r) \equiv \int_0^t n_n(M_r,t') \; v_{\rm th} \; dt'$,
is the neutron exposure 
that would be experienced by a nucleus
if it remained at $M_r$ at all times (TEM00).  No nucleus 
has this history because of convective mixing in the star,
but Fig. \ref{fig:taun} clearly shows where neutrons were liberated
during the star's evolution.

The central neutron exposure of each curve 
in Fig. \ref{fig:taun} is mostly
due to the neutron exposure during core helium burning (TEM00).
Farther out are several peaks of neutron exposures produced
during shell carbon burning. As in the core helium burning, 
the highest neutron exposure occurs
at the innermost convective region
where the temperature and density are the highest
and the neutron-liberating reactions the fastest.
It is worth remembering that convection continually replenishes the
supply of neutron sources to these zones.
The outermost peak of the neutron exposure curves in Fig. \ref{fig:taun}
are due to the shell helium burning. The width of the peak is
due to the outward migration of the innermost part of the helium
shell during its evolution.

While the neutron exposure as function of interior radius, $\tau_n(M_r)$
is a good tool to show where nuclei are exposed to
neutrons, it is less effective in showing the degree of
production of heavy nuclei.  As mentioned above, convective mixing
moves nuclei into and out of the regions of high neutron density, so
no nucleus actually experiences an exposure $\tau_n(M_r)$.
A direct measure of the global production 
of heavy elements is $N_c = \int n_c(M_r) \; dM_r$, 
the number of neutron captures per iron seed
nucleus at different phases of the stellar evolution integrated over
mass range above the mass cut, 1.5 $M_{\sun}$ and
within the relevant burning zone.
$n_c(M_r)$ is the number of neutron captures per iron seed nuclei
at interior radius $M_r$ at the burning phase.
In Table \ref{tab:sprocmeasure} we present $N_c$ 
of our stellar models. 
That table shows that in each stellar model,
the s-process during core helium burning
is the dominant producer of ejected s-process heavy elements,
followed by
the s-process during carbon burning, and then the s-process
during shell helium burning, except for the case of 15 $M_{\sun}$ 
where the mass cut of 1.5 $M_{\sun}$ is comparable to the
maximum size of its convective helium burning core of 2.22 $M_{\sun}$.

If we compare $N_c$ of different the stellar masses 
in Table \ref{tab:sprocmeasure}, we find 
that the larger the stellar mass, 
the greater the heavy element production 
in each s-process burning phase. 
It is interesting to note that our 30 $M_{\sun}$ stellar model
produces a larger total amount of heavy elements than the 
25 $M_{\sun}$ stellar model even after weighting by an initial
mass function factor. We surmise that the largest 
weak s-process production is for stellar mass around 25 - 30 $M_{\sun}$.

\section{ Comparison of Stellar Yields with Solar Abundance}
\label{sec:SolarComparison}

Stellar nucleosynthesis yields can be tested with abundance measurements
of interstellar medium, stellar atmosphere, presolar grains
in meteorites, or solar system abundance.
In this section, we compare our stellar yields with solar system abundance.
In order to make a proper comparison with solar system abundances, 
a Galactic Chemical Evolution calculations 
where time-integrated yield contributions from multi generations of stars
of different metallicities  should be carried
out \citep{1999ARA&A..37..239B}.  However, since
we only produce a limited mass range of stellar models and only of 
initial solar metallicity, a meaningful comparison is done in a 
simplistic approach that 
we compare the s-only solar abundances with the sum of
the IMF-averaged yields of 
the s-only nuclei of our stellar models
and of the s-only nuclei contribution from the main component.
We use the overproduction factors, X/X$_{\sun}$ for the comparisons.
As the overproduction factors of the main component are
the ratios of the values of
the third and the second columns of Table 2 of \citet{1999ApJ...525..886A}.
Our approach is similar with the analysis performed by
\citet{1999ApJ...525..886A} in decomposing the solar abundance distribution
into the s- and r-process components. 
\citet{1999ApJ...525..886A} calculated the r-component residuals 
by subtracting the s abundances of the arithmetic average of 
their 1.5 and 3 $M_{\sun}$ models at Z= Z$_{\sun}/2$
from the solar abundance.
They showed in their Fig. 3 that with their new (n,$\gamma$) cross sections
of neutron magic nuclei at N=82, the agreement between their
low-mass asymptotic giant branch s-only nuclei yields
and their corresponding solar abundances
improved significantly. 

The IMF-averaged overproduction factors of our stellar models, 
y$_{wk}$ is calculated as
 y$_{wk}$ = [r$_{12.5-17.5}$ $\times$ y$_{15}$ $\times$ 13.5 +
             r$_{17.5-22.5}$ $\times$ y$_{20}$ $\times$ 18.5 +
             r$_{22.5-27.5}$ $\times$ y$_{25}$ $\times$ 23.5 +
             r$_{27.5-40.0}$ $\times$ y$_{30}$ $\times$ 28.5 ]
            /(r$_{12.5-17.5}$ $\times$ 13.5 + r$_{17.5-22.5}$ $\times$ 18.5 + 
              r$_{22.5-27.5}$ $\times$ 23.5 + r$_{27.5-40.0}$ $\times$ 28.5)
where we assume each stellar model ejects all its material into interstellar
medium except for its 1.5 $M_{\sun}$ remnant and only stars in the mass
range of 12.5 $M_{\sun}$ and 40 $M_{\sun}$ eject weak s-process materials.
y$_{15}$, y$_{20}$, y$_{25}$, and y$_{30}$ are 
the stellar overproduction factors of our 
15, 20, 25, and 30 $M_{\sun}$ models 
(some are listed in Table \ref{tab:spyield}). 
The factors
r$_{12.5-17.5}$, r$_{17.5-22.5}$, r$_{22.5-27.5}$, and r$_{27.5-40.0}$ are
the normalized-number of stars  in the mass range of 
12.5-17.5 $M_{\sun}$,  17.5-22.5 $M_{\sun}$, 
22.5-27.5 $M_{\sun}$, and 27.5-40.0 $M_{\sun}$,
respectively, assuming their ratios follow
the IMF distribution $\xi_0 m^{-\alpha}$
with Salpeter's original value $\alpha$=1.35.

Our weak s-only nuclei overproduction factors, y$_{wk}$, 
are scaled and summed with
the scaled main s-only nuclei overproduction factors to produce 
the total s-only nuclei overproduction factors, y$_{s-tot}$ such that:
y$_{s-tot}$ = c$_{wk}$ $\times$y$_{wk}$ + c$_{mn}$ $\times$y$_{mn}$.
The scale factors c$_{wk}$ and c$_{mn}$ are determined by 
least-square fit of the total s-only overproduction factors 
relative to the solar values of unity.
We use the 34 s-only nuclei from $^{70}$Ge up to $^{208}$Pb in the fit.
Their standard deviations are taken from column 4 of Table 2 of
\citet{1999ApJ...525..886A} where the values were determined by taking
into account the uncertainties in cross sections and solar abundances.
The best fit of the total s-only nuclei overproduction factors
is represented by the solid circles in Fig. \ref{fig:Cshelln0fit}.
In this figure, the s-only  overproduction factors of the weak component 
are from our stellar models (15N, 20N, 25N, and 30N)
at the end of core oxygen burning 
and are represented by the filled diamond symbols,
whereas the s-only overproduction factors of the main component 
are represented by the solid squares. 
The value of the best-fit $\chi^2$  is 153.7 with 32 degrees of freedom. 
The value of the best-fit $\chi^2$ is quite large suggesting that we 
may be underestimating the standard deviations
(we have not taken
into account the error propagation of the cross sections to the yields
of our stellar models). 
Alternatively, the large $\chi^2$ may be due to our too simple
treatment of chemical evolution or the fact our stellar models
begain with initial solar composition.
Nevertheless, we show that the fit of the s-only nuclei of the weak component 
(A $<$ 90) is as good a fit as the fit of the main component (A $>$ 90).
We expect that if the overproductions due to explosive burning and 
from the complete set of nuclei from the
main components are included, most of the points 
on Fig. \ref{fig:Cshelln0fit} would lie near the dashed line 
in the figure (solar values).

Table \ref{tab:imfcomparesolar} presents the best fit overproduction factors
of the weak and main components
of the s-only nuclei,
of nuclei in the mass range 60 $<$ A $<$ 90 and overproduction factor $>$0.5,
and of some other interesting heavy nuclei.
We find that our set of stellar models using NACRE reaction rates
produces too many $^{70}$Ge and $^{82}$Kr nuclei,
at maximum the excesses are 14\% and 13\%, respectively.
For s-only nuclei with A$\leq$87, 
the weak component contributes at least 40\% of
the solar s-only nuclei.
For s-only nuclei with A$>$90, most of the weak component contributions are
between 5.3\% and 8.5\% except for $^{152}$Gd, $^{187}$Os, and $^{198}$Hg
which are 14\%, 13\%, and 11\%, respectively. 

In principle, the method presented in this section 
is similar to the classical approach of fitting $\sigma N_s$ curve 
solar abundance
pioneered by \citet{1965ApJS...11..121S} and \citet{1967ApJ...148...69C}.
Both methods fit the s-only nuclei of solar abundances.
In the classical approach, seed nuclei are exposed with three exponential
distributions of neutron exposures 
(the main, the weak, and the strong component), 
whereas in the stellar models
seed nuclei are exposed with a single exposure in massive stars 
\citep{1986ana..work..375B,1989ApJ...339..962B}
and an exponential exposure from repeated thermal pulses 
in the low-mass AGB stars \citep{1973exnu.conf..139U}.

\subsection{Comparison of s-process Burning Phase and Single Stellar Model}
\label{sec:CompareSingle}

It is interesting to know how good the s-only nuclei overproduction factors
of each stellar model at the end of core helium burning, at the end of core
oxygen burning, and their IMF-averaged are relative to the best overall fit 
to the s-only solar abundance distribution.
In Fig. \ref{fig:Overprodfits} 
we present the best-fit overproduction factor
distribution of model 30N 
at the end of core helium burning (panel a),
of the IMF-averaged of our stellar models 
at the end of core helium burning (panel b),
of model 25N at the end of core oxygen burning (to include s-process results
from core helium and shell carbon burnings, panel c),
and of the IMF-averaged of our stellar models 
at the end of core oxygen burning (panel d).
In panel d, instead of model 25N for 
the 25 $M_{\sun}$  contribution (shown in Fig.  \ref{fig:Cshelln0fit}), 
we take  model 25K for testing the effect of 
$^{12}$C($\alpha$,$\gamma$)$^{16}$O reaction in 
fitting the s-only solar abundance. 

Model 30N (Fig. \ref{fig:Overprodfits}a)
produces the best s-only solar distribution fit 
with $\chi^2$=176
among 15N, 20N, 25N, 25K, and 30N models 
for yields at the end of core helium burning.
Model 30N also produces a better fit than 
the IMF-averaged of models at the end of
core helium burning ($\chi^2$=205, Fig. \ref{fig:Overprodfits}b),
mostly due to the large
$\chi^2$ contribution from the 15 and 20  $M_{\sun}$ models.

Model 25K (Fig. \ref{fig:Overprodfits}c)
produces the best fit to the s-only solar distribution
with $\chi^2$=161
among 15N, 20N, 25N, 25K, and 30N models 
for yields at the end of core oxygen burning.
We also find the overproduction of s-only nuclei of 
the IMF-averaged of models at the end of core oxygen burning
($\chi^2$=153)
gives a better fit to the s-only solar distribution than
the overproductions of the
IMF-averaged of models at the end of core helium burning
($\chi^2$=206).
The differences in $\chi^2$ between the fits in the panels of
Fig. \ref{fig:Overprodfits}  being larger than 8
are quite significant since
the differences only involve 6 data points for the weak component.
We conclude that
including shell carbon burning s-processing indeed
gives a better fit to the s-only solar distribution nuclei
than using yields from the core helium burning s-process only.
Also IMF-averaging is necessary to give a better fit to the solar abundance
distribution (as can be seen by comparing the $\chi^2$ and the
spread from min to max
of overproduction factors 
of panel c with
the distribution of panel  d).
Furthermore, the overproduction factors X/X$_{\sun}$ $>$ 0.5
for nuclei with 60 $\leq$ A $\leq$ 90 
suggest that solar abundances of nuclei in this
mass range are dominantly produced by the s-processing in massive stars
(see also Table \protect\ref{tab:imfcomparesolar}).

\section{Summary and Discussion of the Results of the s-process}
\label{sec:summary}
Tables \ref{tab:spresults}, \ref{tab:spsummary}, and \ref{tab:spyield} 
summarize our results on
the s-process in the massive stars under consideration. 
In Table \ref{tab:spresults}, 
we emphasize the following points:
\begin{itemize}
 \item A comparison between the overabundance obtained at 
       the end of core He-burning
       in 25N and 25K shows that the reaction 
       $^{12}{\rm C}(\alpha,\gamma)^{16}{\rm O}$
       has only a small influence on the efficiency of 
       the s-process during this phase.
       In contrast, the efficiency of the s-process during 
       shell C-burning is very sensitive
       to the mass fraction of carbon left over at 
       the end of core He-burning, which is
       determined by this reaction.
 \item The overabundances we have obtained in our case 25C at the end of core
       He-burning are similar to those calculated by 
       \citet{1991ApJ...367..228R}.
       but our results of the shell C-burning are different from 
       those by Raiteri et al. (1991b),
       since they have done essentially a one-zone calculation at 
       fixed temperature and density. 
 \item Table \ref{tab:spsummary} indicates that the s-process 
       during core He-burning
       leads to a monotonic increase of the overabundance as 
       a function of stellar mass. 
       However, this does not apply in the case of shell C-burning
       because branchings along the
       s-process path become effective as a result of 
       the higher neutron density encountered
       during this phase (see Fig. \ref{fig:ndin25n}). 
       The behavior of the overabundance of
       $^{63,65}$Cu, $^{64}$Zn,
       $^{80}$Kr, $^{86,87}$Sr, and $^{152}$Gd indicates this feature.
 \item In Table \ref{tab:spyield}, we summarize the stellar yield 
       compared to solar of the
       listed nuclei as obtained by integration above 1.5 $M_{\sun}$ 
       for each stellar mass.
       Their overproduction factor distribution as function of 
       mass number A are plotted in Fig. \ref{fig:syield}.
       The dependence of this yield on the stellar mass reflects 
       the behavior of the overabundance
       described above. 
       Relatively high yield is obtained for the pure s-nuclei.
       We emphasize the remarkable difference by a factor 4.2 in 
       the yield for $^{80}$Kr resulting from
       the sequences 25N and 25K at the end of core oxygen burning. 
       The reason is the
       destruction of $^{80}$Kr by shell C-burning in the sequence 25N 
       and its production in 25K.
       This is seen in Fig. \ref{fig:finalx}, 
       where the normalized mass fractions are displayed
       as a function of the interior mass in
       the sequences 25N and 25K at
       the end of core oxygen burning. 
       Note also the difference in the mass fraction of $^{70}$Zn
       between the two sequences, which we have attributed to 
       the different neutron densities
       encountered during shell C-burning as discussed above 
       in \S \ref{sec:spcshell}.
 \item It is quite remarkable that $^{152}$Gd, produced
       abundantly (overabundance $>$18)
       during core helium burning in massive stars, is
       brought back to its solar value at the end of shell carbon burning
       due to the larger
       $^{152}$Eu(n,$\gamma$) rate during shell carbon burning,
       which causes the s-process flow to bypass $^{152}$Gd.
       This result is reasonable since s-processing in 
       thermally-pulsed AGB stars produces enough $^{152}$Gd 
       to account for its solar abundance \citep{1993ApJ...419..207R}.
       A similar case also occurs for $^{158}$Dy, which has an
       overabundance value of larger than 10 in all models studied 
       at the end of core helium burning \citep{2000A&A...354..740R} but
       decreases to an overabundance of less than solar after
       shell carbon burning. Production of $^{158}$Dy occurs in the
       s-process because $^{157}$Gd, which is stable in the lab, can
       $\beta^-$ decay in stars.  The rate for this decay is temperature
       and density dependent \citep{1987ADNDT..36..375T}.  Interestingly,
       this rate is lower in the conditions of shell carbon burning
       than is core helium burning.  Moreover, the neutron-capture rate
       for $^{157}$Gd increases with the higher temperature and density
       of the carbon shell.  These effects cause the s-process flow to
       bypass $^{158}$Gd during carbon burning.
 \item The opposite case of $^{152}$Gd is for isotope $^{116}$Cd,
       which is destroyed almost completely during core helium burning 
       to an overabundance of less than 0.002, but then reproduced to
       a value close to solar after shell carbon burning due to
       a higher neutron-capture rate of $^{115}$Cd during shell carbon burning.
       A case similar to but less dramatic than $^{116}$Cd is $^{96}$Zr,
       which is destroyed during core helium burning to an overabundance
       $\leq$0.4 then recovers to an overabundance $\geq$1 at the end of
       shell carbon burning.
 \item An interesting feature of the s-process in shell carbon burning
       is the strong enhancement of $^{80}$Se, $^{86}$Kr, and $^{87}$Rb
       to an overabundance larger than 10 from a value of $\sim$1
       at the end of core helium burning (see also
       \citealp{1993ApJ...419..207R}).
 \item Another significant feature of shell carbon burning is
       the overabundances of $^{23}$Na and $^{27}$Al.
       The overabundance of $^{23}$Na is less than 10 at the end of core
       helium burning.  It is enhanced significantly to a value larger
       than 230 in all models studied here.  
       A similar result is also obtained for $^{27}$Al overabundance
       which rises from 1.5 to 75.

\end{itemize}

\section{Conclusions}
\label{sec:conlude}
\begin{figure}
\epsscale{0.7}
   \rotatebox{0}{\includegraphics[height=6.5in]{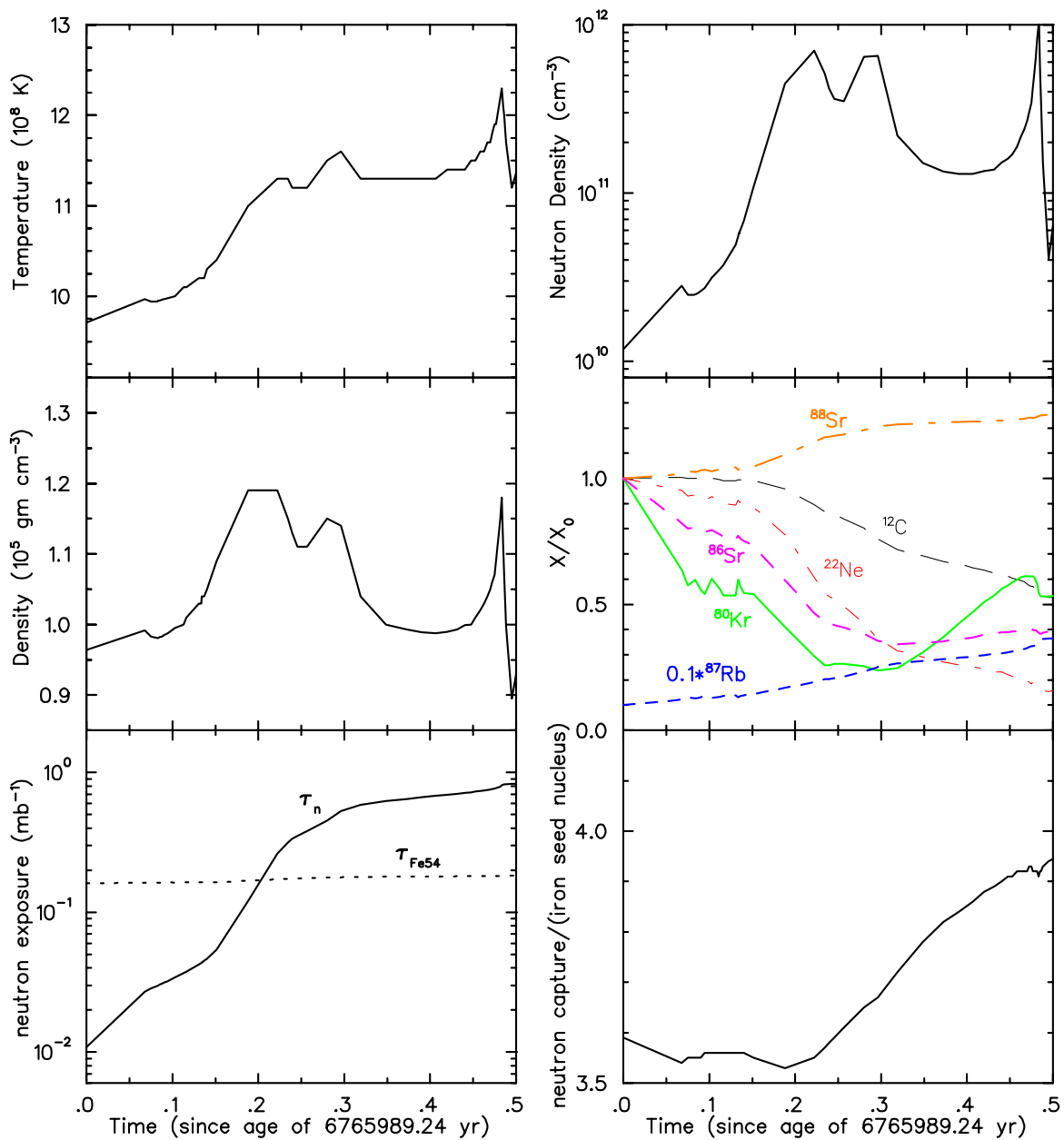}} 
   \caption{ Several physical variables characterizing the carbon-burning
   shell in the sequence 25N. The panels display snapshots taken at 
   mass coordinate
   $M_{\rm r}$=2.26 $M_{\sun}$, which locates the bottom of 
   the carbon-burning shell in this sequence of models.
   Note the gradual increase of the neutron density following the
   gradual change of temperature and density.
   X$_0$ is the mass fraction at the beginning of the shell C-burning.
   $\tau_n$ is the neutron exposure of the shell coordinate
   seen by a nucleus if it stays at this position at all times.
   $\tau_{Fe54}$ is the neutron exposure implied by the mass fraction of
   $^{54}$Fe: $\tau_{Fe54}$ = -ln(X$_{54}$/X$_{54}^0$)/$\sigma_{T}$
   where $\sigma_{T}$ is the neutron-capture cross section at T=30 keV
   and X$_{54}$, X$_{54}^0$ are the final and initial mass fraction
   of $^{54}$Fe, respectively.  $\tau_{Fe54}$ is useful as a measure
   of the neutron exposure averaged over the convective zones.
\label{fig:ndin25n}
   }
\end{figure}

\begin{figure}
\epsscale{0.7}
 \rotatebox{0}{\includegraphics[height=6.5in]{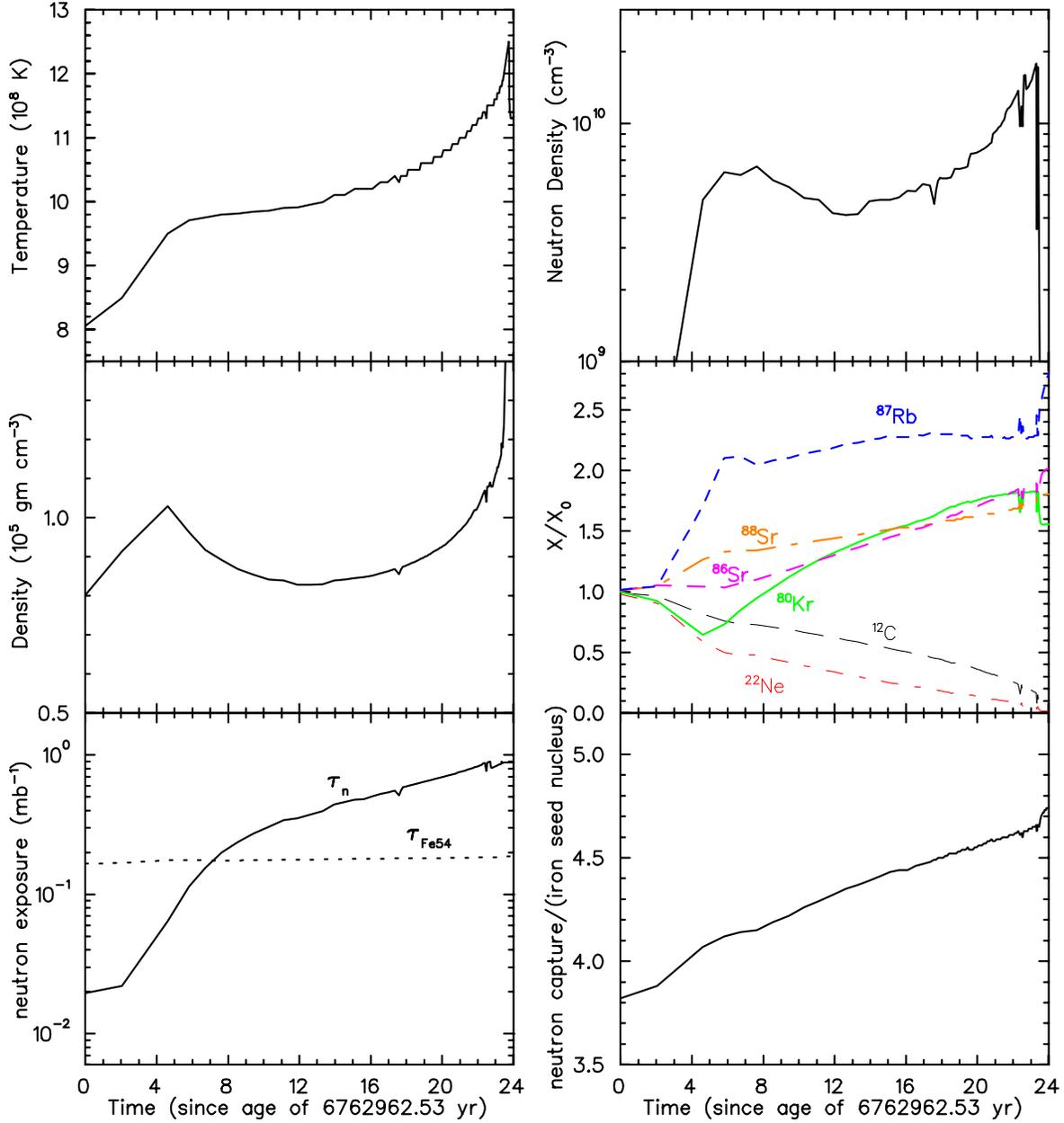}} 
\caption{ The same as Fig. \ref{fig:ndin25n} for the sequence 25K.
 The quantities are taken at a mass coordinate $M_{\rm r}$=1.38 $M_{\sun}$,
 which is the location of the bottom of the carbon-burning shell in this
 model. Note the gradual increase of the neutron density
 following the  gradual change of temperature and density.
 \label{fig:ndin25k}
 }
\end{figure}

\begin{figure}
\epsscale{0.7}
\includegraphics[angle=90,height=2.25in]{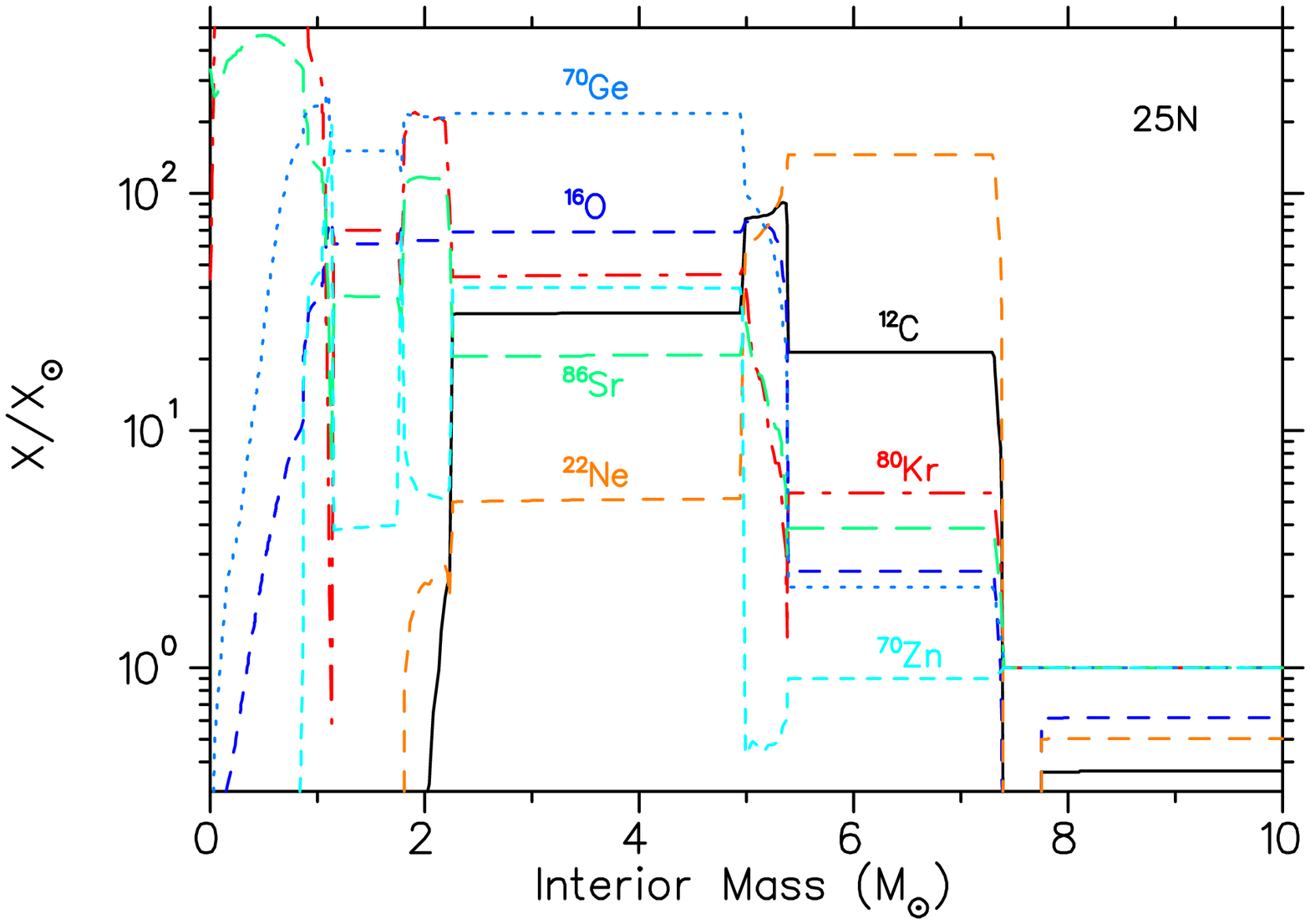}
\includegraphics[angle=90,height=2.25in]{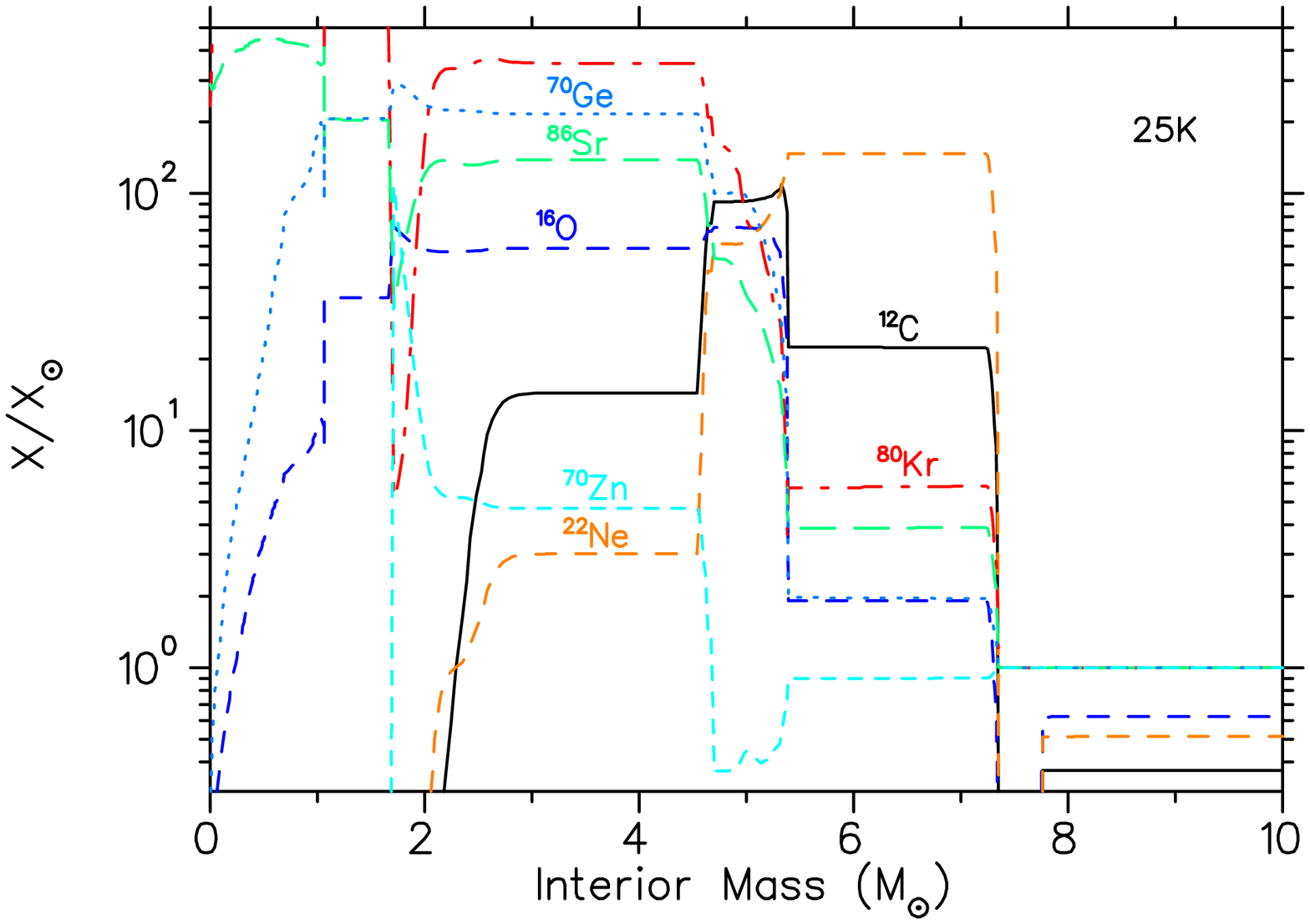}
 \caption{Mass fractions of various important nuclear species
          normalized to solar values versus interior mass.
          These curves represent snapshots at the end of oxygen burning 
          in a 25 $M_{\sun}$ star for model sequences 25N and 25K.
 \label{fig:finalx}
   }
\end{figure}

\begin{figure}
\epsscale{0.7}
 \rotatebox{90}{\includegraphics[height=6.5in]{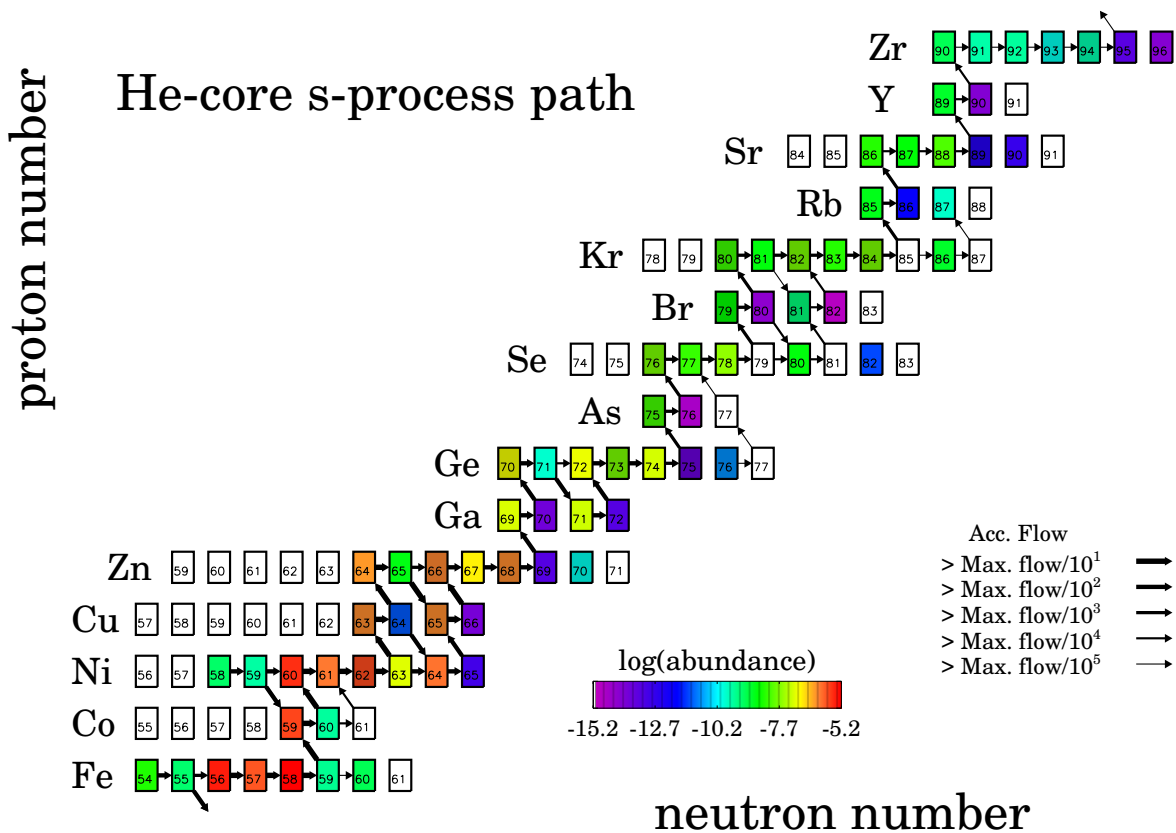}} 
 \caption{s-process nuclear reaction flow and final abundance
   for the s-process during core
   He-burning in a one-zone nucleosynthesis calculation
   using the central temperature, density, and $^{4}$He mass fraction
   tracks of our evolutionary sequence 25N.
   The thickness of an arrow shows the level of that reaction flow
   (i.e. $\sum_n$ N$_A<\sigma v> \rho$ y$_i$ y$_j$ dt$_n$) relative to
   the maximum reaction flow within the boundary of the chart.
   The largest neutron-capture flows within the range of the figure
   are the $^{56,57,58}$Fe(n,$\gamma$) reactions.
 \label{fig:hepath}
  }
\end{figure}

\begin{figure}
   \epsscale{0.7}
 \rotatebox{90}{\includegraphics[height=6.5in]{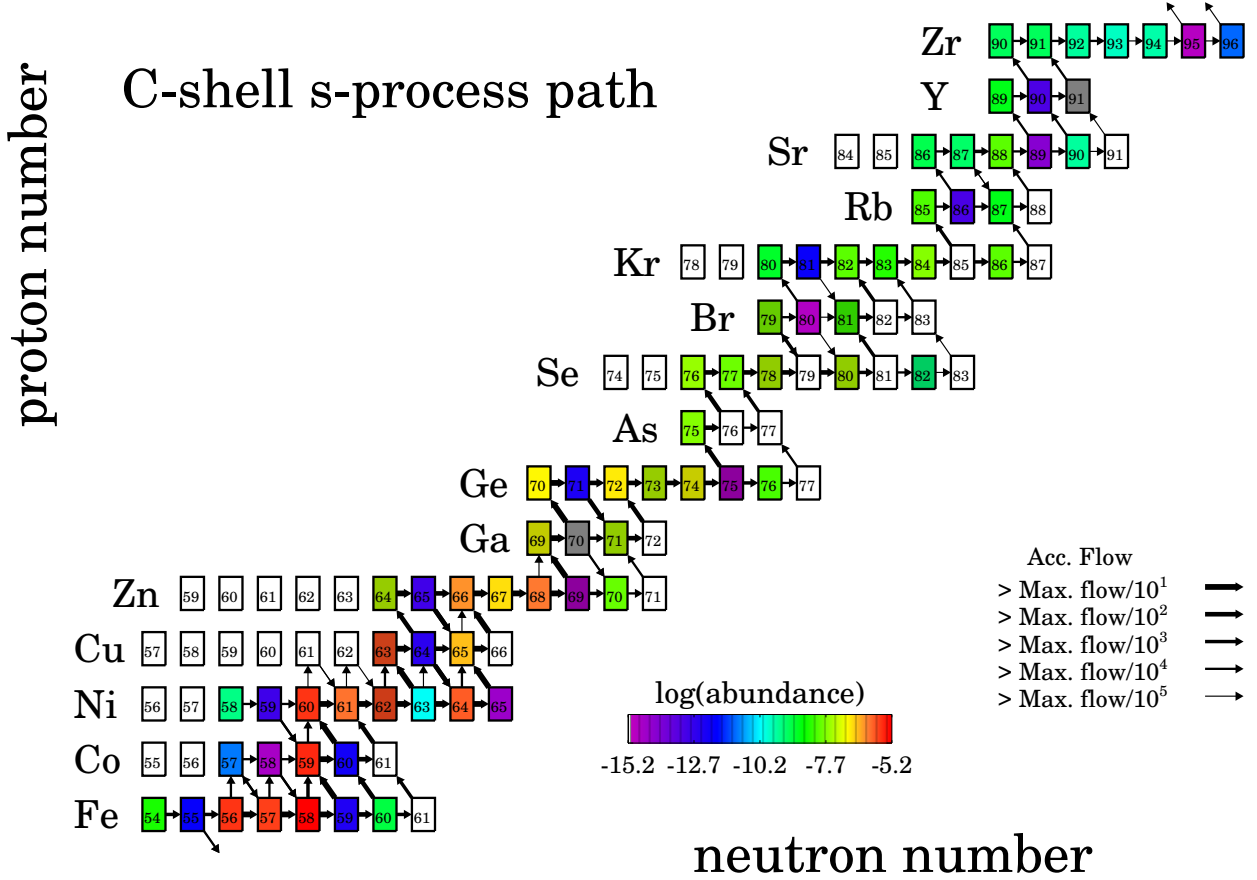}} 
 \caption{Chart of nuclear reaction flow for the s-process during
  shell carbon burning in one-zone calculation
  using the temperature, density, and $^{12}$C mass fraction tracks
  of the innermost shell of convective C-burning 
  of our evolutionary sequence 25N.
  The largest neutron-capture flows within the range of the figure
  are $^{58}$Fe(n,$\gamma$), $^{57}$Fe(n,$\gamma$), and $^{56}$Fe(n,$\gamma$)
  reactions.  The largest proton-capture flows within the range of the figure
  are the $^{58}$Fe(p,g), $^{57}$Fe(p,g), and $^{56}$Fe(p,g)  reactions.
  The largest neutron-producer reactions are the
  $^{22}$Ne($\alpha$,n), $^{21}$Ne($\alpha$,n), $^{17}$O($\alpha$,n),
  $^{13}$C($\alpha$,n), and $^{26}$Mg($\alpha$,n).
  The ratios of their relative strengths are 
  0.60:0.19:0.15:0.03:0.03, respectively.
 \label{fig:cpath}
}
\end{figure}

\begin{figure}
 \epsscale{0.7}
 \rotatebox{90}{\includegraphics[height=5.0in]{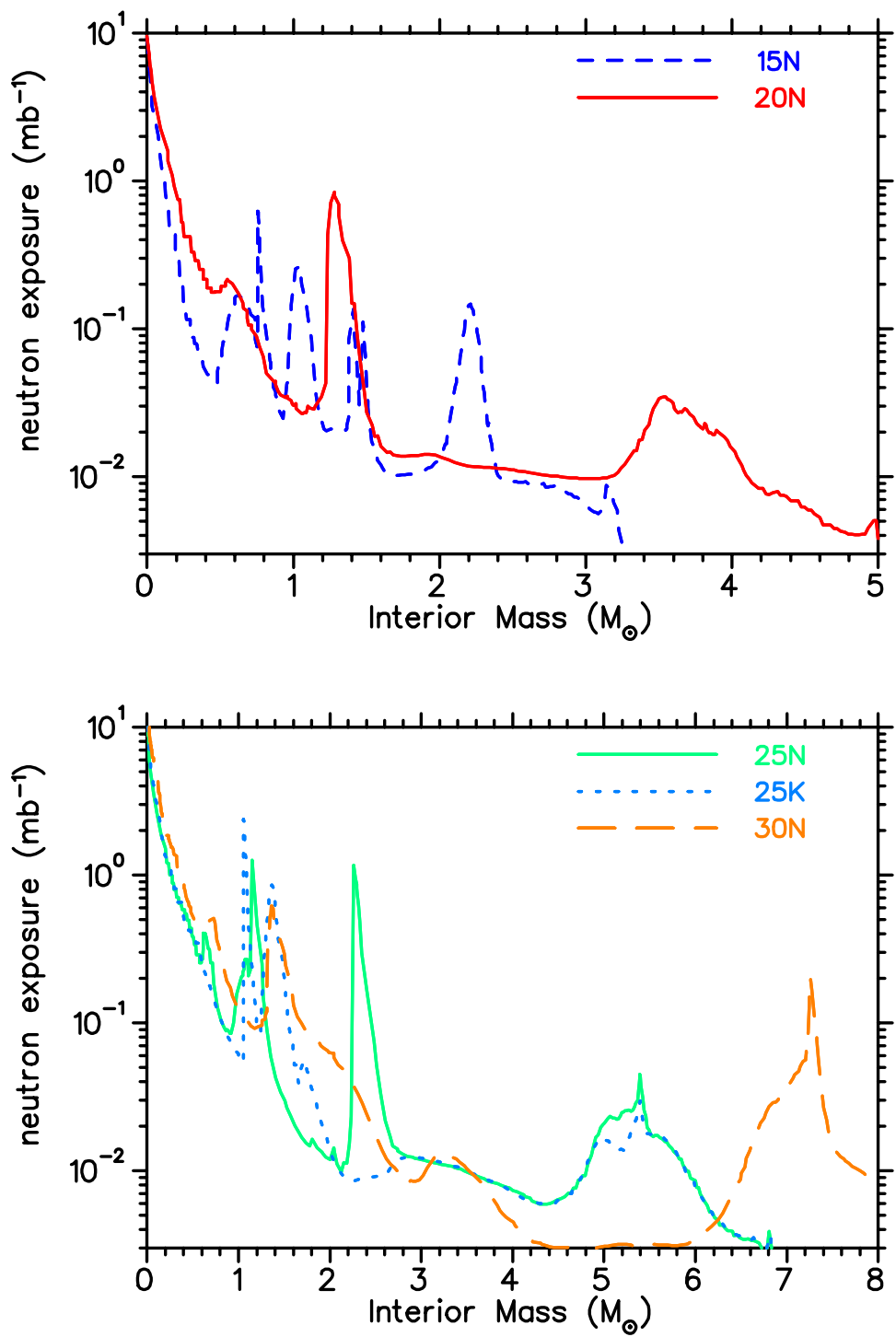}} 
 \caption{Neutron exposure versus interior mass for 
          sequences 15N, 20N, 25N, 25K, and 30N 
          taken at the end of core oxygen burning. 
          The curves indicate the history and the location of s-process nuclear
          burning in the stellar models. 
          The baselines of the curves are due to the s-processing during
          core helium burning with their highest values at the
          central region of the models.
          The narrow peaks superimposed on the generally falling curve
          arise from neutron
          exposure during different phases of shell carbon burning, 
          except the outermost
          broad peak, which is due to the neutron exposure in
          the helium-burning shell 
          (see also Table \ref{tab:shellcarbon}).
 \label{fig:taun}
}
\end{figure}

\begin{figure}
 \epsscale{0.7}
 \rotatebox{90}{\includegraphics[height=6.5in]{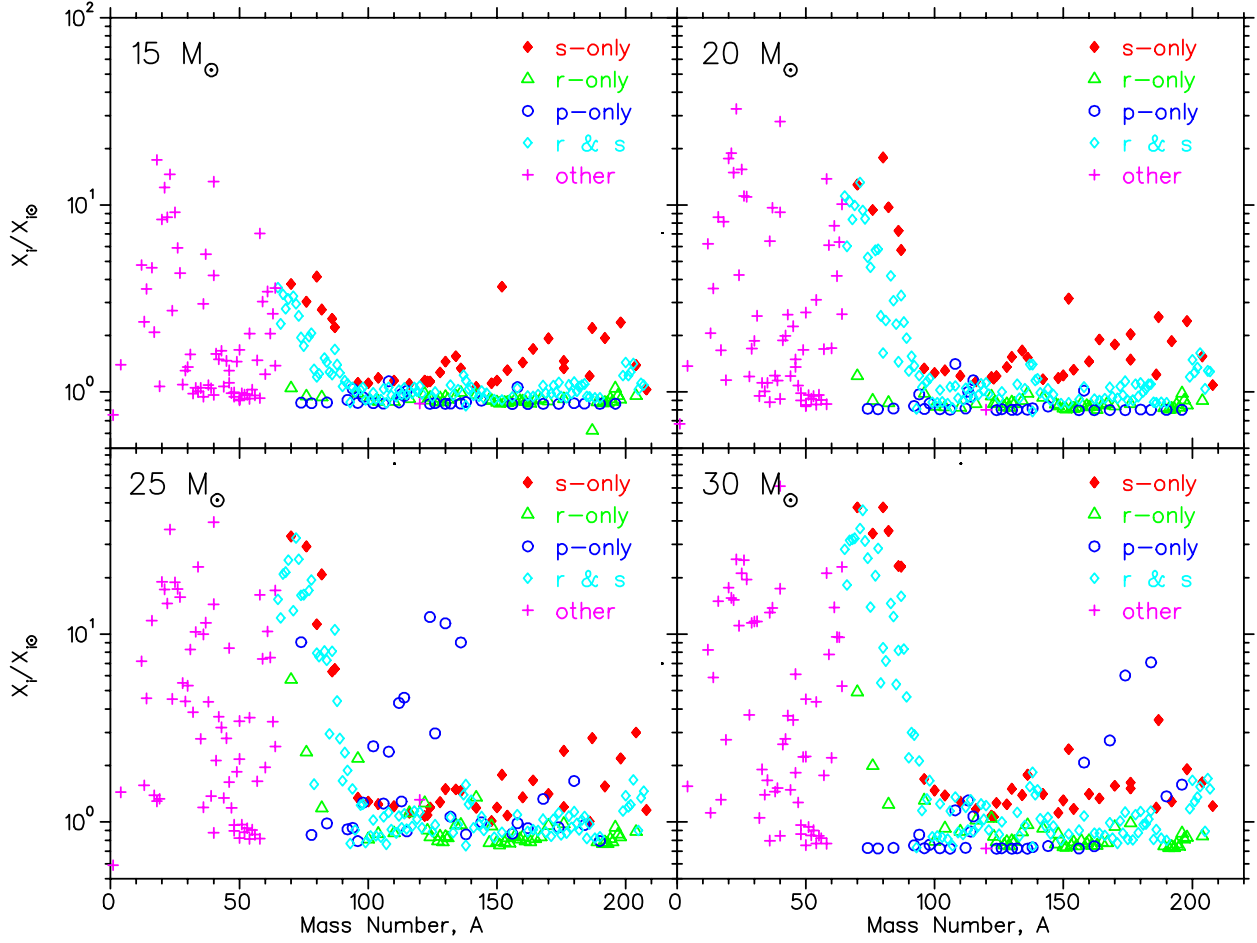}} 
 \caption{
The overproduction factor distribution averaged over the ejecta of
models 15N, 20N, 25N, and 30N.
 \label{fig:syield}
}
\end{figure}

\begin{figure}
 \epsscale{0.7}
 \rotatebox{90}{\includegraphics[height=6.0in]{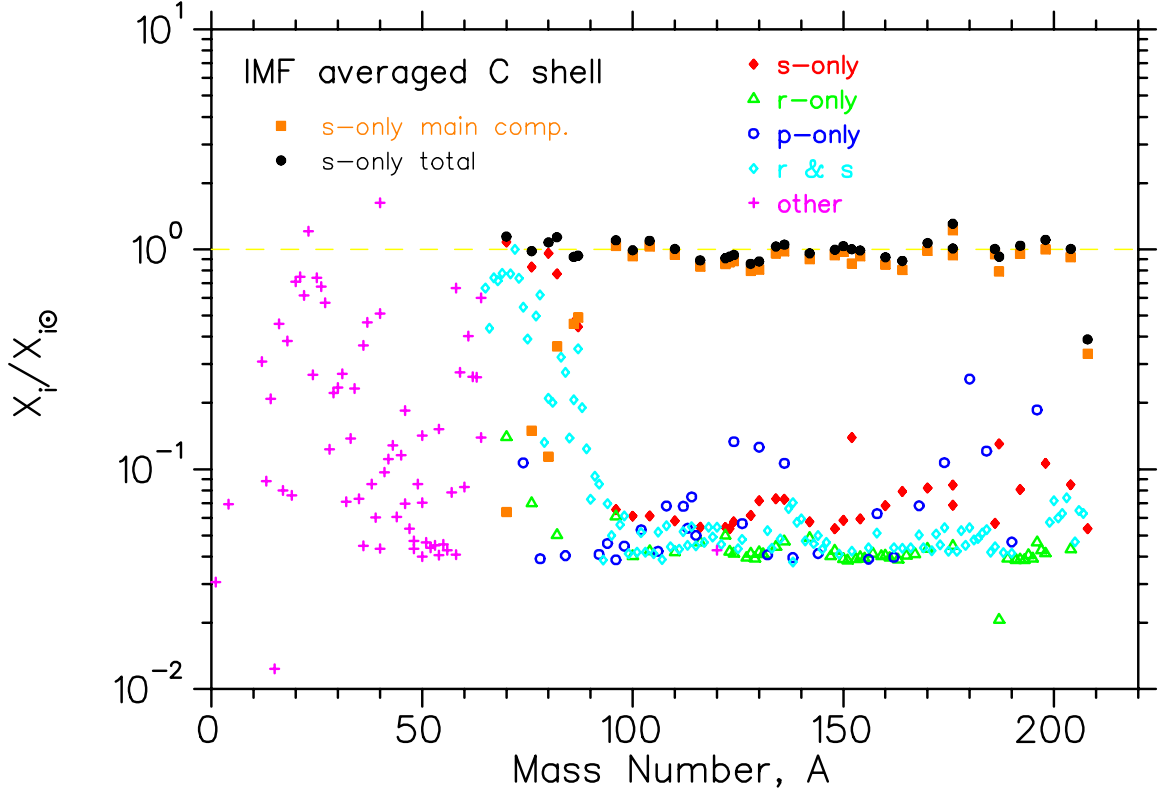}} 
 \caption{
   The overproduction factor distribution of the IMF-averaged of
models 15N, 20N, 25N, and 30N that gives best fit to the s-only nuclei
solar distribution after adding the s-only from the main component to
the weak component.  
The $\chi^2$ of the best fit is 153 with 32 degrees of freedom.
The primary nucleosynthesis production process for each nuclei is 
indicated by the symbol type.
The s-only contribution from the main component is shown with solid 
squares.
The s-only contribution from the weak component is shown with solid
diamonds and the total s-only abundance
of the weak and main component is shown
with solid circles. The dashed line represents the solar abundance
distribution.
Note that we only include the s-only nuclei of
the main component in the plot.
 \label{fig:Cshelln0fit}
}
\end{figure}

\begin{figure}
 \epsscale{0.7}
 \rotatebox{90}{\includegraphics[height=6.5in]{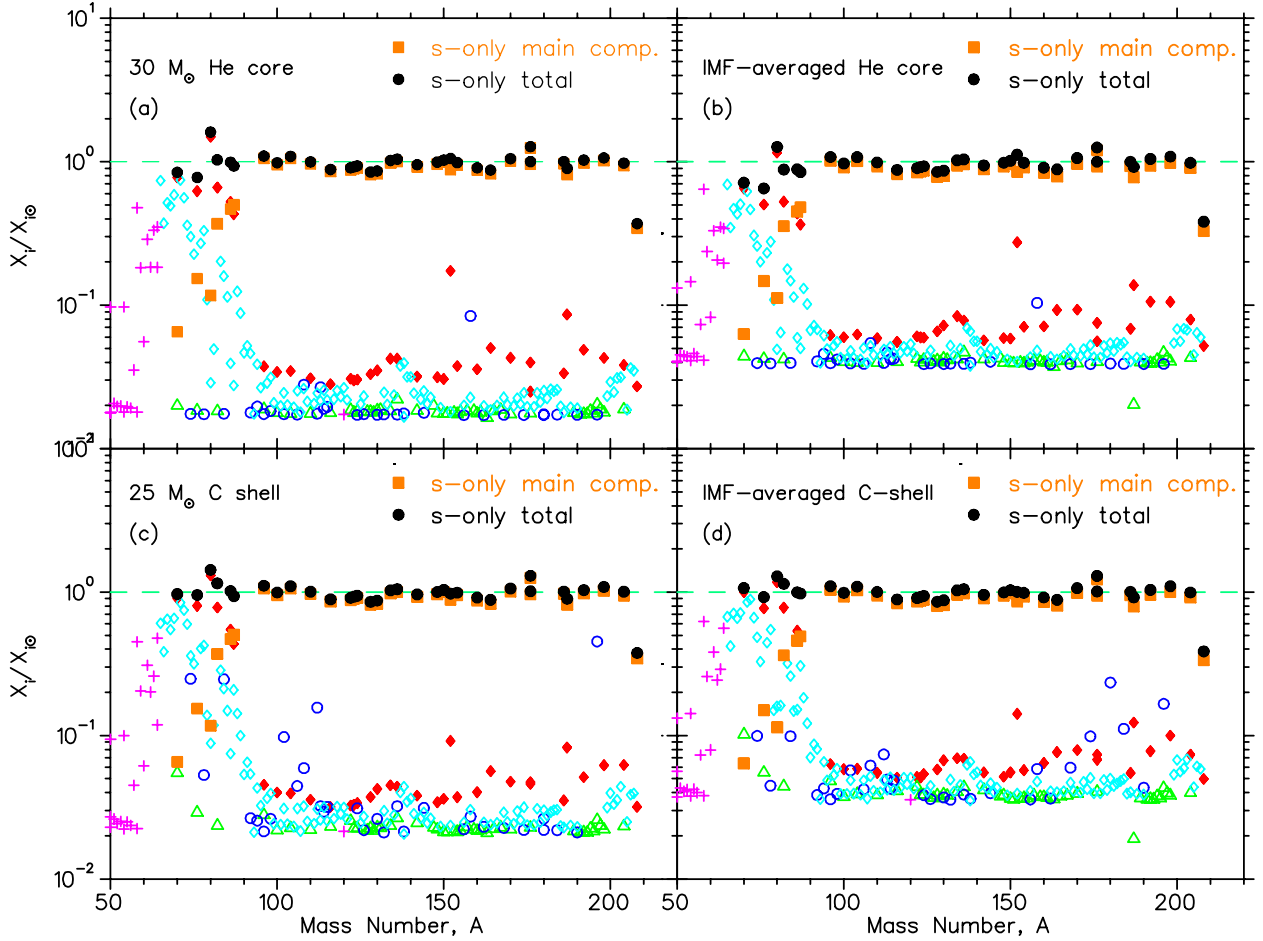}} 
 \caption{Similar to Fig. \protect\ref{fig:Cshelln0fit} but for 
model 30N at the end of core helium burning 
($\chi^2$=176 see text, panel a: top left),
for the IMF-averaged of models 15N, 20N, 25N, and 30N at the end of
core helium burning ($\chi^2$=205, panel b: top right),
for model 25K at the end of core oxygen burning 
($\chi^2$=161, panel c: bottom left),
and for the IMF-averaged of models 15N, 20N, 25K, and 30N 
at the end of core oxygen burning ($\chi^2$=153, panel d: bottom right).
The improvement of the fits going from 
s-only product of the core helium burning
to the product of shell carbon burning shows the importance of including
s-process nucleosynthesis from shell carbon burning in fitting the
solar abundance.
Also averaging the overproduction factors over the stellar mass range
is necessary to fit the solar abundance distribution.
The small spread of overproduction factor, X/X$_{\sun}$ $>$ 0.5
for nuclei with 60 $\leq$ A $\leq$ 90 
suggests that solar abundance nuclei in this
mass range are dominantly produced by the s-processing in massive stars
(see also Table \protect\ref{tab:imfcomparesolar}).
 \label{fig:Overprodfits}
}
\end{figure}

Our detailed study of the s-process nucleosynthesis resulting 
from core He-burning
and shell C-burning in massive stars on the basis of the updated 
nuclear data of
some relevant reactions reveal many interesting points 
which we summarize in the following.
\begin{itemize}
 \item The efficiency of s-process nucleosynthesis during 
       core He-burning does not depend
       on the rate of 
       $^{12}{\rm C}(\alpha,\gamma)^{16}{\rm O}$ 
       but it is sensitive to 
       the rates of $^{22}{\rm Ne}(\alpha,{\rm n})^{25}{\rm Mg}$ and
       $^{16}{\rm O}({\rm n},\gamma)^{17}{\rm O}$. 
       When we use the updated rates of these two reactions,
       as described in \S \ref{sec:smodels}, 
       we find a significantly reduced efficiency of the s-process
       during core He-burning (see Table \ref{tab:hesprocess}).

 \item The s-process in shell C-burning is more complicated and 
       depends on the evolution of
       the massive star beyond core He-burning.
       This complexity can be seen from
       the locations, the number of carbon convective shells, and
       the thickness of carbon convective shells in our models.
       These in turn are sensitive to the central carbon mass fraction
       X$_{12}$ achieved at the end of core He-burning as a result 
       of the reaction
       $^{12}{\rm C}(\alpha,\gamma)^{16}{\rm O}$, whose
       rate is not yet adequately determined.
       If this rate leads to
       a relatively lower X$_{12}$, then the s-process occurs 
       later in time, possibly even after central neon burning.
       Consequently, the neutrons density achieved is high enough 
       (see Fig. \ref{fig:ndin25n}) to drive the
       nuclear reaction flow to the Zirconium region. 
       This does not happen when X$_{12}$ is higher, as
       in our case 25K, since the s-process occurs here earlier 
       in time (before central neon ignition)
       and at lower temperatures and densities, which result
       in a smaller neutron density.
       This explains why the nuclear reaction flow stops 
       essentially in the Strontium region.

 \item Our calculations show that the overabundance of $^{70}$Zn 
       can be used as indicator of the
       strength of the nuclear reaction flow through 
       the branchings along the s-process path,
       especially at $^{69}$Zn.
       We have also found that $^{87}$Rb is strongly produced 
       during shell carbon burning
       due to the higher rate of neutron-capture of $^{86}$Rb 
       relative to its rate during core helium burning
       \citep{1993ApJ...419..207R}.

 \item We measure the s-processing in the core helium, shell helium,
       and shell carbon burning in massive stars with 
       $N_c = \int n_c(M_r) \; dM_r$ and show their relative 
       strengths or importance. 
       We show the s-process contribution from
       shell carbon burning decreases with increasing mass of the star.

 \item In comparing the yields of s-only nuclei of our stellar models with
the solar distribution, we find that it is necessary to include 
the results of s-processing from shell carbon burning and 
to mix the yields of all mass range of massive stars to give
a reasonable fit to the solar distribution.
For s-only nuclei with mass number A$\leq$87, massive stars contribute
at least 40\% to the solar s-only nuclei. 
For s-only nuclei with mass number A$>$90, massive stars contribute
on average $\sim$7\%, except for 
$^{152}$Gd, $^{187}$Os, and $^{198}$Hg
which can be 14\%, 13\%, and 11\%, respectively. 

\end{itemize}
\label{sec:conclusion}

\acknowledgments
The authors are grateful to J. S. Brown for providing the equation of
state tables.
M.F. EL Eid thanks the American University of Beirut (AUB) for
a URB grant to visit Clemson University during summer 2005.
L.-S. The thanks the Aspen Institute of Physics for hospitality,
support, and for organizing the 2005
workshop on the physics of s-process.
L.-S. The also acknowledges valuable discussions with 
A. Burrows, D. Clayton, I. Dominguez, 
R. Gallino, A. Heger, F. Herwig, 
F. Kappeler, U. Ott, R. Reifarth, O. Straniero, and F.X.Timmes.
This work has been supported by NSF grant AST-9819877
and by grants from NASA's Cosmochemistry Program
and from the DOE's Scientific Discovery
through Advanced Computing Program (grant DE-FC02-01ER41189).




\bibliographystyle{apj}

\clearpage

\begin{deluxetable}{cccc}
\footnotesize 
\tablecolumns{4} 
\tablecaption{
   Rates Used for Important Reactions in the Calculations}
\tablewidth{0pc} 
\tablehead{ \colhead{Model} & 
            \colhead{$^{12}{\rm C}(\alpha,\gamma)^{16}{\rm O}$} &
            \colhead{$^{16}{\rm O}({\rm n},\gamma)^{17}{\rm O}$} &
            \colhead{$^{22}{\rm Ne}(\alpha,{\rm n})^{25}{\rm Mg}$}
          }
\startdata
15N     &  NACRE   & \citet{1995ApJ...441L..89I} & NACRE \\
20N     &  NACRE   & \citet{1995ApJ...441L..89I} & NACRE \\
25C     &  CF88    & \citet{1992ApJS...80..403B} & CF88 \\
25K     &  Kunz et al. (2002) & \citet{1995ApJ...441L..89I} & NACRE \\
25N     &  NACRE   &  \citet{1995ApJ...441L..89I} & NACRE \\
30N     &  NACRE   &  \citet{1995ApJ...441L..89I} & NACRE \\
TEM00   &  CF88    &  BVW            & CF88 \\
\enddata
 \label{tab:modelrate}
\end{deluxetable}

\begin{deluxetable}{lcccccc}
\footnotesize \tablecolumns{7} \tablecaption{
   Comparison of s-Processing in Massive Stars during Core Helium Burning
   among different authors}
\tablewidth{0pc} \tablehead{ \colhead{Author} & \colhead{$\tau_c$} &
                   \colhead{n$_{\rm c}^a$} & \colhead{$\langle\tau\rangle$}  &
                   \colhead{$n_{\rm n}^{max}$} &
                   \colhead{X$_{22}$}  &
                   \colhead{X$_{80}$/X$_{80 \odot}$$^e$} \\
\colhead{}       &  &
                   \colhead{}  & \colhead{(mb$^{-1}$)$^b$}  &
                   \colhead{($ \times 10^5$ cm$^{-3}$)$^c$} &
                   \colhead{($ \times 10^{-2}$)$^d$} & \colhead{}
}
\startdata
\multicolumn{7}{c}{\underline{\textbf{15} $M_{\sun}$} } \\
K94                       & --   & 1.80  &  0.09  & 2.05  & 1.65 & 21  \\
TEM00,A                   & 4.00 & 3.38  &  0.10  & 2.27  & 1.33 & 117 \\
Present Work, (15N)       & 1.79 & 1.19  &  0.06  & 0.89  & 1.50 & 15  \\

\multicolumn{7}{c}{\underline{\textbf{20} $M_{\sun}$} } \\
K94                       & --   & 3.66  &  0.15  & 5.06  & 1.32 & 116  \\
TEM00,A                   & 5.93 & 5.48  &  0.16  & 3.50  & 1.04 & 598  \\
Present Work, (20N)       & 3.38 & 2.34  &  0.10  & 1.63  & 1.12 & 56   \\

\multicolumn{7}{c}{\underline{\textbf{25} $M_{\sun}$} } \\
K94                    & --   & 5.41  &  0.20  & 6.62  & 1.00 & 475  \\
TEM00,A                & 7.15 & 6.70  &  0.22  & 4.24  & 0.76 & 1100 \\
Present Work, (25C)    & 5.43 & 5.14  &  0.30  & 1.95  & 0.98 & 618 \\
Present Work, (25N)    & 4.81 & 3.52  &  0.15  & 2.60  & 0.77 & 174 \\
Present Work, (25K)    & 5.00 & 3.63  &  0.15  & 2.53  & 0.78 & 186 \\
Present Work, (25NM)   & 5.19 & 4.03  &  0.17  & 2.50  & 0.58 & 264 \\

\multicolumn{7}{c}{\underline{\textbf{30} $M_{\sun}$} } \\
K94                    & --   & 6.55  &  0.23  & 6.74  & 0.79 & 933  \\
TEM00,A                & 8.09 & 7.63  &  0.22  & 4.44  & 0.65 & 1368 \\
Present Work, (30N)    & 5.94 & 4.24  &  0.14  & 1.84  & 0.54 & 352  \\
\enddata

\tablenotetext{a}{Number of neutrons captured per iron seed averaged
over the maximum convective core mass.}
 \tablenotetext{b}{Mean neutron exposure at 30 keV
  averaged over the convective core mass.}
\tablenotetext{c}{Maximum of the mean neutron density.}
\tablenotetext{d}{Final $^{22}$Ne mass fraction.}
\tablenotetext{e}{Final $^{80}$Kr production factor averaged over
the maximum convective core mass.}
 \label{tab:hesprocess}
\end{deluxetable}

\begin{deluxetable}{cccccc|cccc}
\tabletypesize{\footnotesize}
 \tablecolumns{10} \tablewidth{0pc}
 \tablecaption{Overabundance factors (X/X$_\bigodot$) resulting
 from the s-process calculation of the present work compared to
 \citet{1991ApJ...367..228R, 1991ApJ...371..665R},
  referred to as R(91b) and R(91a)  }
 \tablehead{
             \colhead{species} & \colhead{Z} &
             \multicolumn{4}{c}{End Core Helium} &
             \multicolumn{4}{c}{C-Shell} \\
             \cline{3-6} \cline{7-10} 
             \colhead{} & \colhead{} &  
             \colhead{25N} & \colhead{25K} & \colhead{25C} & \colhead{R(91b)} &
            \colhead{25N} & \colhead{25K} & \colhead{25C} & \colhead{R(91a)}  
 }
 \startdata
 $^{23}$Na & 11 &  6.97 &  7.01 &  7.70 &\nodata & 235&  315   &  240  & \nodata  \\
 $^{27}$Al & 13 &  1.16 &  1.15 &  0.91 & \nodata & 103 &  143 &  95.4 & \nodata  \\
 $^{37}$Cl & 17 &  72.1 &  72   &  72.5 &  65.8&  61.1&  61.8  &  62   & \nodata  \\
 $^{40}$K  & 19 & 268   & 284   & 332   & 291.7& 224  & 255    & 280   & \nodata  \\
 $^{50}$Ti & 22 & 158   & 158   &  19.7 &  15.9&  16.9&  16.4  &  21.4 & \nodata  \\
 $^{54}$Cr & 24 &  16.8 &  15.8 &  16.2 &  16.5&  16.4&  16.4  &  15.3 & \nodata   \\
 $^{58}$Fe & 26 & 105   & 104   &  90   &  84  &  92.8&  93.1  &  76.1 & 56.7 \\
 $^{59}$Co & 27 &  36.4 &  36.5 &  34   &  35.9&  39.2&  45.5  &  33.3 &  \nodata  \\
 $^{61}$Ni & 28 &  53.8 &  55.2 &  61   &  84.6&  60.4&  67.9  &  68   & \nodata  \\
 $^{62}$Ni & 28 &  31.6 &  32.7 &  40   &  49.9&  34.4&  36.6  &  40.2 &  \nodata \\
 $^{64}$Ni & 28 &  56.6 &  59.1 &  95.6 & 164.5& 109  & 115    & 153   &  \nodata \\
 $^{63}$Cu & 29 &  58.3 &  60.8 &  78.2 &  91.8&  11.5&  64.6  &  42.9 &  \nodata  \\
 $^{65}$Cu & 29 & 122   & 128   & 205   & 226.3&  83.5 & 148   & 175   &  \nodata   \\
 $^{64}$Zn & 30 &  29.4 &  30.7 &  43.6 &  41  &   8.6 &  23.2 &  22.5 &  \nodata   \\
 $^{66}$Zn & 30 &  57   &  59.6 & 107   & 118.9&  62.1 &  80.6 & 121   &  \nodata   \\
 $^{67}$Zn & 30 &  79.4 &  82.9 & 153   & 171.7& 137   & 160   & 256   &   \nodata  \\
 $^{68}$Zn & 30 &  70   &  73.1 & 158   & 164.7& 128   & 131   & 237   &  \nodata   \\
 $^{70}$Zn & 30 &  0.38 &   0.38&   0.56&  --  &  39.9 &   4.7 & 31.7  &  \nodata   \\
 $^{70}$Ge & 32 & 107   & 112   & 270   & 253.7& 217   & 216   & 402   & 527.1\\
 $^{72}$Ge & 32 &  72.2 &  75.2 & 201   & 190.7& 187   & 158   & 385   & \nodata   \\
 $^{73}$Ge & 32 &  45.1 &  46.9 & 128   & 128.8& 180   & 147   & 357   &  \nodata  \\
 $^{74}$Ge & 32 &  35.9 &  37.5 & 110   &  99.3& 101   &  84.9 & 204   &  \nodata  \\
 $^{75}$As & 33 &  26.3 &  27.4 &  81.9 &  59.6&  99.1 &  75.4 & 189   &  \nodata   \\
 $^{76}$Se & 34 &  74.7 &  78.2 & 241   & 212  & 189   & 164   & 357   & 763.1\\
 $^{80}$Se & 34 &  1.21 &  1.29 & 3.95  & \nodata & 54.7 & 56.3  & \nodata & 763.1\\
 $^{80}$Kr & 36 & 174   & 183   & 618   & 480.7&  45.2 & 354   & 367   & 675.6\\
 $^{82}$Kr & 36 &  73.4 &  77.9 & 277   & 210.3& 124   & 181   & 485   & 495.9\\
 $^{86}$Kr & 36 &   2.57&   2.63&   5.72& \nodata &  50.9 &  10.4 &  44.3 & 224.3\\
 $^{87}$Rb & 37 &   1.26&   1.27&   3.03& \nodata &  55.4 &  43.3 & 233   & 292.3\\
 $^{86}$Sr & 38 &  57.1 &  60.7 & 232   & 147  &  20.7 & 138   & 316   & 147.4\\
 $^{87}$Sr & 38 &  47.3 &  50.4 & 190   & 129  &  28.3 & 108   & 378   &  57.3\\
 $^{88}$Sr & 38 &  14.1 &  14.9 & 45.3  & \nodata  &  20.1 & 27.7  & 99.6  & \nodata \\
 $^{96}$Zr & 40 &   0.19&   0.19&   0.12& \nodata &  11.3 &   1.18&   3.87& \nodata  \\
$^{116}$Cd & 48 & 0.001 & 0.001 & 0.001 & \nodata & 2.67 & 0.34&   4.44& \nodata \\
$^{152}$Gd & 64 &  22.8 &  22.9 &  31.8 &  38.6&   0.09&   8.24&   1.19&  29.2\\
$^{158}$Dy & 66 &  14.9 &  11.5 &  21.4 & \nodata & 0.04&  0.38&   0.87& \nodata \\
\enddata
\label{tab:spresults}
\end{deluxetable}

\begin{deluxetable}{cr|rrrr|rrrr}
\tabletypesize{\scriptsize}
\tablecolumns{10} \tablewidth{0pc} 
\tablecaption{Overabundance of some relevant isotopes in He-Core
and C-shell of our stellar models.} 
\tablehead{ 
\colhead{Isotope} & \colhead{Z} &
\multicolumn{4}{c}{Core Helium Burning}  & 
\multicolumn{4}{c}{Carbon Shell Burning}  \\
\cline{3-6} \cline{7-10}
\colhead{} & \colhead{} &  
\colhead{15N}   &  \colhead{20N}    & \colhead{25N}     & \colhead{30N} &   
\colhead{15N}   & \colhead{20N}     & \colhead{25N}     & \colhead{30N } 
} 
\startdata
$^{23}$Na & 11   &     6.14  &    6.77  &    6.97  &    6.98   &     311.92  &   342.2  &  235.02  &  240.13 \\
$^{27}$Al & 13   &     0.89  &    0.99  &    1.16  &    1.31   &      83.61  &  110.03  &  103.11  &  126.31 \\
$^{37}$Cl & 17   &    58.78  &   69.29  &   72.12  &   72.69   &      57.65  &   62.22  &   61.14  &   58.47 \\
$^{40}$K  & 19   &   154.13  &  216.89  &  267.58  &  299.45   &     146.56  &  190.70  &  224.21  &  277.73 \\
$^{50}$Ti & 22   &     9.78  &   13.63  &   15.84  &   17.42   &      11.55  &   14.81  &   16.89  &   18.16 \\
$^{54}$Cr & 24   &    13.03  &   16.15  &   16.84  &   16.76   &      15.09  &   16.74  &   16.45  &   16.15 \\
$^{58}$Fe & 26   &    82.03  &  106.33  &  104.89  &   98.54   &      98.23  &  102.28  &   92.78  &   88.76 \\
$^{59}$Co & 27   &    19.68  &   32.95  &   36.43  &   36.18   &      39.66  &   46.76  &   39.25  &   40.23 \\
$^{61}$Ni & 28   &    19.96  &   39.98  &   53.82  &   59.59   &      39.54  &   61.78  &   60.39  &   68.44 \\
$^{62}$Ni & 28   &     9.53  &   21.10  &   31.55  &   37.34   &      17.74  &   30.25  &   34.42  &   38.32 \\
$^{64}$Ni & 28   &    11.10  &   31.27  &   56.63  &   76.70   &      48.56  &   95.50  &  109.32  &  116.65 \\
$^{63}$Cu & 29   &    16.16  &   37.42  &   58.32  &   70.70   &      33.12  &   48.55  &   11.48  &   45.89 \\
$^{65}$Cu & 29   &    23.37  &   67.45  &  122.41  &  165.68   &      42.18  &   95.37  &   83.51  &  158.52 \\
$^{64}$Zn & 30   &     6.79  &   17.44  &   29.40  &   37.72   &       5.47  &   14.51  &    8.61  &   21.91 \\
$^{66}$Zn & 30   &     8.45  &   28.59  &   56.97  &   81.94   &      21.17  &   49.73  &   62.10  &   96.00 \\
$^{67}$Zn & 30   &    10.85  &   38.69  &   79.37  &  116.15   &      40.82  &   96.97  &  136.82  &  197.35 \\
$^{68}$Zn & 30   &     7.42  &   30.19  &   70.13  &  111.04   &      29.29  &   76.69  &  128.42  &  161.20 \\
$^{70}$Zn & 30   &     0.51  &    0.38  &    0.38  &    0.41   &       2.65  &    3.87  &   39.94  &    6.16 \\
$^{70}$Ge & 32   &     9.83  &   42.61  &  107.36  &  178.42   &      44.89  &  124.58  &  217.37  &  270.10 \\
$^{72}$Ge & 32   &     5.96  &   25.98  &   72.22  &  128.22   &      32.49  &   89.61  &  186.75  &  207.79 \\
$^{73}$Ge & 32   &     3.59  &   15.86  &   45.12  &   81.42   &      29.93  &   82.63  &  179.56  &  196.67 \\
$^{74}$Ge & 32   &     2.88  &   12.09  &   35.86  &   67.27   &      16.50  &   46.90  &  100.75  &  109.58 \\
$^{75}$As & 33   &     2.09  &    8.77  &   26.29  &   49.76   &      14.58  &   40.43  &   99.09  &   96.53 \\
$^{76}$Se & 34   &     6.01  &   24.58  &   74.75  &  143.74   &      31.08  &   89.80  &  188.51  &  209.01 \\
$^{80}$Se & 34   &     0.15  &    0.39  &    1.21  &    2.51   &       4.75  &    8.09  &   54.71  &   27.70 \\
$^{80}$Kr & 36   &    15.24  &   55.70  &  174.25  &  352.13   &      49.58  &  182.40  &   45.16  &  316.64 \\
$^{82}$Kr & 36   &     7.83  &   23.59  &   73.44  &  153.31   &      26.29  &   93.21  &  123.66  &  204.95 \\
$^{86}$Kr & 36   &     0.94  &    1.34  &    2.57  &    4.92   &       3.71  &    5.88  &   50.92  &   16.33 \\
$^{87}$Rb & 37   &     0.72  &    0.84  &    1.26  &    2.21   &      17.83  &   27.23  &   55.45  &   76.23 \\
$^{86}$Sr & 38   &    11.99  &   22.61  &   57.07  &  121.10   &      24.79  &   61.24  &   20.71  &  152.34 \\
$^{87}$Sr & 38   &    11.14  &   20.28  &   47.27  &   98.75   &      22.00  &   49.36  &   28.32  &  150.39 \\
$^{88}$Sr & 38   &     3.70  &    7.53  &   14.11  &   25.13   &       7.89  &   14.45  &   20.07  &   40.14 \\
$^{96}$Zr & 40   &     0.42  &    0.27  &    0.20  &    0.16   &       1.12  &    1.14  &   11.30  &    1.93 \\
$^{116}$Cd & 48  &     0.0002&    0.001 &    0.001 &    0.002  &       0.55  &    0.30  &    2.67  &    0.47 \\
$^{152}$Gd & 64  &    17.65  &   18.81  &   22.76  &   27.04   &       2.38  &    3.74  &    0.09  &    0.91 \\
$^{158}$Dy & 66  &     7.52  &   10.90  &   14.91  &   18.71   &       0.35  &    0.14  &    0.04  &    0.64 \\
\enddata
\label{tab:spsummary}
\end{deluxetable}

\begin{deluxetable}{cccccccccccc}
\tabletypesize{\scriptsize}
 \tablecolumns{12} \tablewidth{0pc}
 \tablecaption{Properties of s-processing during Shell Carbon Burning}
 \tablehead{
  \colhead{Model} & \colhead{$\Delta$ $M_{\rm C-Shell}^a$} &
  \colhead{$\tau_{\rm CB}^b$} & \colhead{T$_9^c$} & \colhead{$\rho_5^d$} &
  \colhead{$\Delta \tau_n^e$} & \colhead{$\Delta$ n$_c^f$} & \colhead{$<\Delta\tau>^g$} &
  \colhead{n$_{10}^{max \; h}$} & \colhead{X$_{22}^i$} & 
  \colhead{$\mathcal{O}_{80}^j$} & \colhead{$\mathcal{O}_{88}^{j,k}$} 
 \\
     & \colhead{($M_{\sun}$)} & \colhead{(yrs)}  &  & 
                              & \colhead{(mb$^{-1}$)} &           &
       \colhead{(mb$^{-1}$)} & \colhead{} & 
       \colhead{($\times$10$^{-3}$)}  &  &  
 }
 \startdata
15N  & 1.38 - 1.98 &  1.07 & 1.07 & 1.81 & 0.20 & 1.04 & 0.038 & 3.9 & 3.10 & 28 & 7.59 \\
20N  & 1.24 - 3.12 & 21.1  & 0.97 & 1.10 & 0.71 & 1.22 & 0.049 & 1.0 & 0.88 & 177 & 14.0 \\
25C  & 1.23 - 4.00 & 12.0  & 1.01 & 0.98 & 0.80 & 1.32 & 0.042 & 1.3 & 1.06 & 911 & 97.4 \\
25K  & 1.30 - 4.54 & 20.3  & 1.02 & 0.91 & 0.83 & 1.02 & 0.036 & 1.1 & 0.37 & 352 & 27.7 \\
25N  & 2.26 - 4.94 & 0.46  & 1.15 & 1.14 & 1.10 & 0.47 & 0.038 & 70  & 0.67 & 45  & 20.1 \\
25NM & 1.21 - 3.50 & 4.12  & 1.08 & 1.30 & 0.40 & 1.14 & 0.079 & 1.8 & 0.81 & 108 & 44.7 \\
30N  & 1.33 - 5.22 & 4.06  & 1.13 & 1.10 & 0.56 & 0.61 & 0.019 & 4.3 & 0.80 & 311 & 46.9 \\
\enddata
\tablenotetext{a}{The interior mass range of the last carbon shell.}
\tablenotetext{b}{The duration of the shell carbon burning.}
\tablenotetext{c}{The average temperature weighted by its neutron exposure at the bottom of the carbon shell in 10$^9$ K.}
\tablenotetext{d}{The average density weighted by its neutron exposure at the bottom of the carbon shell in 10$^5$ gm cm$^{-3}$.}
\tablenotetext{e}{The increase of the neutron exposure at the bottom of the carbon shell.}
\tablenotetext{f}{The increase of the number of neutron captures per iron seed averaged over the convective shell.}
\tablenotetext{g}{The increase of the neutron exposure averaged over the convective shell.}
\tablenotetext{h}{The maximum neutron density at the bottom of the carbon shell in 10$^{10}$ cm$^{-3}$.}
\tablenotetext{i}{The mass fraction of $^{22}$Ne at the end of the burning shell.}
\tablenotetext{j}{The overabundance of $^{80}$Kr and $^{88}$Sr isotopes relative to their solar abundance.}
\tablenotetext{k}{For comparison, the overabundances of $^{88}$Sr at the end of core helium burning for model 15N, 20N,
                  25C, 25K, 25N, 25NM, and 30N are 3.46, 7.17, 43.0, 14.4, 13.4, 19.1, and 23.8 respectively.}
\label{tab:shellcarbon}
\end{deluxetable}

\begin{deluxetable}{rrrrrrr}
\tabletypesize{\scriptsize}
\tablecolumns{7} \tablewidth{0pc} \tablecaption{Stellar yield (X/X$_{\bigodot}$ with mass cut
at $M_r$=1.5 $M_{\bigodot}$) of some heavy isotopes at the end of core Oxygen burning} \tablehead{
\colhead{Isotope} & \colhead{Z} & \multicolumn{5}{c}{X/X$_{\bigodot}$} \\
\cline{3-7}
\colhead{} & \colhead{} &
\colhead{15N}   &  \colhead{20N}    & \colhead{25N}     & \colhead{25K}      &   \colhead{30N} 
} \startdata
$^{11}$Na & 11    &   14.60   &    32.61   &    36.20   &    37.61   &    25.01 \\
$^{13}$Al & 13    &    4.33   &    11.12   &    15.82   &    19.81   &    19.61 \\
$^{37}$Cl & 17    &    5.44   &     9.65   &    11.52   &    11.67   &    13.81 \\
$^{40}$K  & 19    &   13.30   &    27.89   &    39.55   &    44.95   &    61.42 \\
$^{50}$Ti & 22    &    1.68   &     2.66   &     3.45   &     3.42   &     4.51 \\
$^{54}$Cr & 24    &    2.06   &     3.10   &     3.60   &     3.64   &     4.37 \\
$^{58}$Fe & 26    &    7.04   &    13.78   &    16.19   &    16.39   &    21.15 \\
$^{59}$Co & 27    &    3.04   &     6.09   &     7.39   &     7.44   &     7.79 \\
$^{61}$Ni & 28    &    3.44   &     7.75   &    10.33   &    11.27   &    13.90 \\
$^{62}$Ni & 28    &    2.05   &     4.17   &     7.52   &     7.32   &     9.65 \\
$^{64}$Ni & 28    &    3.60   &    10.07   &    17.11   &    17.39   &    22.81 \\
$^{63}$Cu & 29    &    2.62   &     6.32   &     3.42   &     9.43   &     9.61 \\
$^{65}$Cu & 29    &    3.61   &    11.14   &    15.31   &    22.05   &    28.26 \\
$^{64}$Zn & 30    &    1.38   &     2.61   &     2.52   &     4.31   &     5.29 \\
$^{66}$Zn & 30    &    2.31   &     6.01   &    12.21   &    13.97   &    18.29 \\
$^{67}$Zn & 30    &    3.32   &    10.40   &    20.88   &    23.59   &    31.57 \\
$^{68}$Zn & 30    &    2.78   &     8.36   &    21.39   &    19.91   &    31.93 \\
$^{70}$Zn & 30    &    1.05   &     1.22   &     5.74   &     2.00   &     4.92 \\
$^{70}$Ge & 32    &    3.77   &    12.85   &    33.15   &    32.93   &    47.42 \\
$^{72}$Ge & 32    &    2.96   &     9.34   &    32.52   &    27.23   &    45.71 \\
$^{73}$Ge & 32    &    2.55   &     8.43   &    25.08   &    21.78   &    31.35 \\
$^{74}$Ge & 32    &    1.95   &     5.25   &    16.12   &    13.13   &    25.41 \\
$^{75}$As & 33    &    1.76   &     4.64   &    16.23   &    11.54   &    13.94 \\
$^{76}$Se & 34    &    3.04   &     9.40   &    29.33   &    29.18   &    34.34 \\
$^{80}$Se & 34    &    1.20   &     1.61   &     7.92   &     3.21   &     8.46 \\
$^{80}$Kr & 36    &    4.14   &    17.93   &    11.29   &    47.72   &    47.35 \\
$^{82}$Kr & 36    &    2.76   &     9.72   &    20.82   &    28.41   &    35.43 \\
$^{86}$Kr & 36    &    1.24   &     1.45   &     8.10   &     2.72   &     8.21 \\
$^{87}$Rb & 37    &    1.69   &     3.28   &    10.55   &     7.58   &    15.93 \\
$^{86}$Sr & 38    &    2.47   &     7.26   &     6.33   &    19.91   &    23.10 \\
$^{87}$Sr & 38    &    2.22   &     5.74   &     6.53   &    15.79   &    22.89 \\
$^{88}$Sr & 38    &    1.40   &     2.36   &     4.40   &     5.18   &     8.33 \\
$^{96}$Zr & 40    &    0.98   &     0.98   &     2.18   &     1.02   &     1.30 \\
$^{116}$Cd & 48   &    0.92   &     0.86   &     1.12   &     0.85   &     1.02 \\
$^{152}$Gd & 64   &    3.66   &     3.15   &     1.78   &     3.34   &     2.44 \\
$^{158}$Dy & 66   &    1.06   &     1.02   &     0.99   &     0.99   &     2.07 \\
\enddata
\label{tab:spyield}
\end{deluxetable}

\begin{deluxetable}{ccccc}
\tablecolumns{5} 
\tablecaption{
  Measuring s-process Production of Core Helium, Shell Helium, and
  Shell Carbon Burning in Massive Stars}
\tablewidth{0pc}
\tablehead{ \colhead{Model} &
            \colhead{$N_c^a$(He-core)} &
            \colhead{$N_c$(He-shell)} &
            \colhead{$N_c$(C-shell)} &
            \colhead{$N_c$(Total)} 
}
\startdata
15N & 0.75 & 0.24 & 0.90 & 1.90 \\
20N & 5.01 & 0.27 & 2.27 & 7.54 \\
25C & 19.3 & 0.54 & 4.96 & 24.8 \\
25K & 13.3 & 0.28 & 3.39 & 17.0 \\
25N & 12.9 & 0.32 & 2.48 & 15.7 \\
25NM &14.5 & 0.36 & 2.28 & 17.1 \\
30N & 23.9 & 0.72 & 4.33 & 29.0 \\
\enddata
\tablenotetext{a}{$N_c = \int n_c(M_r) \; dM_r$, where
$n_c$ is the number of neutron captures per iron seed nuclei, and
integration is over the mass range (in unit of $M_{\sun}$) 
of the relevant convective burning regions
and above the mass cut, 1.5 $M_{\sun}$}
\label{tab:sprocmeasure}
\end{deluxetable}
-----------------------------------------------------------------------

\begin{deluxetable}{ccccccc}
\tabletypesize{\scriptsize}
\tablecolumns{7} \tablewidth{0pc} 
\tablecaption{s-process Contribution to Solar Abundances}
\tablehead{
\colhead{Isotope} & \colhead{Z} & \multicolumn{5}{c}
{Overproduction factor X/X$_{\bigodot}$} \\
\cline{3-7}
\colhead{} & \colhead{} &
\colhead{Weak Comp. (25N)} & \colhead{Weak Comp. (25K)} & \colhead{Main Comp.} &
\colhead{Total (25N)$^a$} & \colhead{Total (25K)$^b$}
} 
\startdata
$^{23}$Na  & 11  &  1.21  &  1.14  &  \nodata  & \nodata & \nodata \\
$^{27}$Al  & 13  &  0.57  &  0.56  &  \nodata  & \nodata & \nodata \\
$^{37}$Cl  & 17  &  0.46  &  0.43  &  \nodata  & \nodata & \nodata \\
$^{40}$K   & 19  &  1.63  &  1.56  &  \nodata  & \nodata & \nodata \\
$^{50}$Ti  & 22  &  0.14  &  0.13  &  \nodata  & \nodata & \nodata \\
$^{54}$Cr  & 24  &  0.15  &  0.14  &  \nodata  & \nodata & \nodata \\
$^{58}$Fe  & 26  &  0.67  &  0.62  &  \nodata  & \nodata & \nodata \\
$^{59}$Co  & 27  &  0.28  &  0.26  &  \nodata  & \nodata & \nodata \\
$^{61}$Ni  & 28  &  0.40  &  0.38  &  \nodata  & \nodata & \nodata \\
$^{62}$Ni  & 28  &  0.26  &  0.24  &  \nodata  & \nodata & \nodata \\
$^{64}$Ni  & 28  &  0.60  &  0.56  &  \nodata  & \nodata & \nodata \\
$^{63}$Cu  & 29  &  0.26  &  0.29  &     0.00  &  0.27   &  0.30 \\
$^{65}$Cu  & 29  &  0.67  &  0.67  &     0.01  &  0.68   &  0.68 \\
$^{64}$Zn  & 30  &  0.14  &  0.14  &     0.00  &  0.14   &  0.14 \\
$^{66}$Zn  & 30  &  0.44  &  0.42  &     0.01  &  0.45   &  0.43 \\
$^{67}$Zn  & 30  &  0.74  &  0.71  &     0.02  &  0.76   &  0.73 \\
$^{68}$Zn  & 30  &  0.72  &  0.66  &     0.03  &  0.75   &  0.69 \\
$^{70}$Zn  & 30  &  0.14  &  0.10  &     0.00  &  0.14   &  0.10 \\
$^{69}$Ga  & 31  &  0.78  &  0.72  &     0.04  &  0.82   &  0.76 \\
$^{71}$Ga  & 31  &  0.77  &  0.85  &     0.06  &  0.82   &  0.90 \\
$^{70}$Ge  & 32  &  1.08  &  1.00  &     0.07  &  1.14   &  1.07 \\
$^{72}$Ge  & 32  &  1.00  &  0.89  &     0.08  &  1.07   &  0.96 \\
$^{73}$Ge  & 32  &  0.74  &  0.66  &     0.05  &  0.79   &  0.71 \\
$^{74}$Ge  & 32  &  0.55  &  0.48  &     0.06  &  0.60   &  0.54 \\
$^{75}$As  & 33  &  0.39  &  0.33  &     0.05  &  0.44   &  0.37 \\
$^{76}$Se  & 34  &  0.83  &  0.77  &     0.15  &  0.98   &  0.92 \\
$^{78}$Se  & 34  &  0.62  &  0.55  &     0.11  &  0.72   &  0.65 \\
$^{80}$Se  & 34  &  0.21  &  0.16  &     0.09  &  0.30   &  0.25 \\
$^{80}$Kr  & 36  &  0.96  &  1.17  &     0.12  &  1.07   &  1.29 \\
$^{82}$Kr  & 36  &  0.77  &  0.78  &     0.37  &  1.13   &  1.14 \\
$^{86}$Kr  & 36  &  0.21  &  0.15  &     0.27  &  0.47   &  0.41 \\
$^{87}$Rb  & 37  &  0.35  &  0.31  &     0.35  &  0.70   &  0.65 \\
$^{86}$Sr  & 38  &  0.47  &  0.54  &     0.47  &  0.92   &  1.00 \\
$^{87}$Sr  & 38  &  0.44  &  0.48  &     0.50  &  0.93   &  0.98 \\
$^{88}$Sr  & 38  &  0.19  &  0.18  &     0.92  &  1.08   &  1.08 \\
$^{96}$Zr  & 40  &  0.06  &  0.05  &     0.55  &  0.60   &  0.58 \\
$^{96}$Mo  & 42  &  0.07  &  0.06  &     1.03  &  1.10   &  1.10 \\
$^{100}$Ru & 44  &  0.06  &  0.06  &     0.95  &  0.99   &  0.99 \\
$^{104}$Pd & 46  &  0.06  &  0.06  &     1.06  &  1.09   &  1.09 \\
$^{110}$Cd & 48  &  0.06  &  0.05  &     0.97  &  1.00   &  1.00 \\
$^{116}$Cd & 48  &  0.05  &  0.04  &     0.18  &  0.22   &  0.21 \\
$^{116}$Sn & 50  &  0.05  &  0.05  &     0.86  &  0.89   &  0.89 \\
$^{122}$Te & 52  &  0.05  &  0.05  &     0.88  &  0.91   &  0.91 \\
$^{123}$Te & 52  &  0.05  &  0.05  &     0.89  &  0.93   &  0.92 \\
$^{124}$Te & 52  &  0.06  &  0.05  &     0.91  &  0.94   &  0.94 \\
$^{128}$Xe & 54  &  0.06  &  0.06  &     0.82  &  0.86   &  0.86 \\
$^{130}$Xe & 54  &  0.07  &  0.07  &     0.83  &  0.88   &  0.88 \\
$^{134}$Ba & 56  &  0.07  &  0.07  &     0.98  &  1.03   &  1.03 \\
$^{136}$Ba & 56  &  0.07  &  0.07  &     1.00  &  1.05   &  1.05 \\
$^{142}$Nd & 60  &  0.06  &  0.06  &     0.92  &  0.96   &  0.96 \\
$^{148}$Sm & 62  &  0.05  &  0.05  &     0.97  &  0.99   &  0.99 \\
$^{150}$Sm & 62  &  0.06  &  0.06  &     1.00  &  1.03   &  1.03 \\
$^{152}$Gd & 64  &  0.14  &  0.14  &     0.88  &  1.00   &  1.00 \\
$^{154}$Gd & 64  &  0.06  &  0.06  &     0.95  &  0.99   &  0.99 \\
$^{158}$Dy & 66  &  0.06  &  0.06  &   \nodata & \nodata & \nodata \\
$^{160}$Dy & 66  &  0.07  &  0.06  &     0.87  &  0.92   &  0.92 \\
$^{164}$Er & 68  &  0.08  &  0.08  &     0.83  &  0.88   &  0.88 \\
$^{170}$Yb & 70  &  0.08  &  0.08  &     1.01  &  1.07   &  1.07 \\
$^{176}$Lu & 71  &  0.08  &  0.07  &     1.25  &  1.30   &  1.30 \\
$^{176}$Hf & 72  &  0.07  &  0.07  &     0.96  &  1.01   &  1.01 \\
$^{186}$Os & 76  &  0.06  &  0.05  &     0.97  &  1.00   &  1.00 \\
$^{187}$Os & 76  &  0.13  &  0.12  &     0.82  &  0.92   &  0.92 \\
$^{192}$Pt & 78  &  0.08  &  0.08  &     0.98  &  1.04   &  1.04 \\
$^{198}$Hg & 80  &  0.11  &  0.10  &     1.02  &  1.10   &  1.10 \\
$^{204}$Pb & 82  &  0.08  &  0.07  &     0.94  &  1.00   &  0.99 \\
$^{208}$Pb & 82  &  0.05  &  0.05  &     0.34  &  0.39   &  0.39 \\
\enddata
\tablenotetext{a}{The values of main component is scaled by a factor of
0.974 in order to produce the best fit to the s-only solar abundance}
\tablenotetext{b}{The values of main component is scaled by a factor of
0.976 in order to produce the best fit to the s-only solar abundance}
\label{tab:imfcomparesolar}
\end{deluxetable}
\clearpage


\clearpage
\clearpage

\clearpage

\clearpage

\clearpage

\clearpage

\clearpage

\clearpage

\clearpage

\clearpage

\clearpage

\clearpage

\clearpage

\clearpage

\clearpage

\clearpage

\end{document}